\documentclass{hotnets15}

\usepackage{times}  
\usepackage{epsfig}
\usepackage[TABBOTCAP]{subfigure}
\usepackage{tabularx}
\usepackage{graphicx} 
\usepackage{color}
\usepackage{xspace}
\usepackage{thumbpdf}
\usepackage{listings}
\usepackage{verbatim}
\usepackage{hyperref}
\usepackage{booktabs}
\usepackage{colortbl}

\hypersetup{pdfstartview=FitH,pdfpagelayout=SinglePage}
\pdfoutput=1
\usepackage{outlines,url,grffile,enumerate,appendix,balance}
\usepackage[normalem]{ulem}
\usepackage{units}
\usepackage{paralist}
\usepackage{amsmath}
\usepackage{xcolor}
\usepackage{changebar}
\usepackage[footnotesize]{caption}
\usepackage{microtype}
\usepackage{cite}
\usepackage{mathptmx}
\usepackage{multirow}
\usepackage{framed}
\usepackage[shortlabels]{enumitem}
\usepackage{enumerate}
\usepackage{etoolbox}
\usepackage{algpseudocode}

\newcommand{\argmin}{\operatornamewithlimits{argmin}}

\makeatletter
\newcommand\subparagraph{%
  \@startsection{subparagraph}{5}
  {\parindent}
  {3.25ex \@plus 1ex \@minus .2ex}
  {-1em}
  {\normalfont\normalsize\bfseries}}
\makeatother
\usepackage[compact]{titlesec}
\let\subparagraph\relax

\usepackage{enumitem}
\setitemize{itemsep=1pt,topsep=1pt,parsep=1pt,partopsep=1pt}

\newcommand{\bi}{\begin{itemize}}
\newcommand{\ei}{\end{itemize}}

\newcommand{\ie}{{\it i.e.,}\xspace}
\newcommand\eat[1]{}
\newcommand\paragraphb[1]{\noindent{\bf #1}}
\newcommand\paragraphi[1]{\noindent\emph{#1}}

\newcommand{\allnotes}[1]{}

\renewcommand{\allnotes}[1]{\textit{#1}}

\newcommand{\fixme}[1]{\allnotes{\bf\textcolor{red}{[#1]}}}
\newcommand{\changed}[1]{#1}

\renewcommand\fixme[1]{}

\def\squarebox#1{\hbox to #1{\hfill\vbox to #1{\vfill}}}

\colorlet{shadecolor}{gray!25}   
{\normalsize \endMakeFramed}

\begin{document}



\title{Universal Packet Scheduling}

\author{\makebox[.05\linewidth]{}Radhika Mittal$^\dag$ \and Rachit Agarwal$^\dag$ \and Sylvia Ratnasamy$^\dag$ \and Scott Shenker$^{\dag\ddag}$\makebox[.05\linewidth]{} \and $^\dag$UC Berkeley \and $^\ddag$ICSI}

\maketitle


\subsection*{Abstract}
In this paper we address a seemingly simple question: {\em Is there a universal packet scheduling algorithm?} More precisely, we analyze (both theoretically and empirically) whether there is a single packet scheduling algorithm that, at a network-wide level, can match the results of {\em any} given scheduling algorithm. We find that in general the answer is ``no''. However, we show theoretically that the classical Least Slack Time First (LSTF) scheduling algorithm comes closest to being universal and demonstrate empirically that LSTF can closely, though not perfectly, replay a wide range of scheduling algorithms in realistic network settings. We then evaluate whether LSTF can be used {\em in practice} to meet various network-wide objectives by looking at three popular performance metrics (mean FCT, tail packet delays, and fairness); we find that LSTF performs comparable to the state-of-the-art for each of them. 


\section{Introduction}

There is a large and active research literature on novel packet scheduling algorithms, from simple schemes such as priority scheduling~\cite{diffserv}, to complicated mechanisms to achieve fairness~\cite{drr,fq,gps}, to schemes that help reduce tail latency~\cite{fifoplus} or flow completion time~\cite{pfabric}, and this short list barely scratches the surface of past and current work. In this paper we do not add to this impressive collection of algorithms, but instead ask if there is a single {\em universal} packet scheduling algorithm that could obviate the need for new ones. \eat{In this context, we consider a packet scheduling algorithm to be both how packets are served inside the network (based on their time of arrival and their packet header) and how packet header fields are initialized at the edge of the network; this definition includes all the classical scheduling algorithms (FIFO, LIFO, priority, round-robin) as well as algorithms that incorporate dynamic packet state~\cite{dps,csfq,fifoplus}. Our definition does not include how routers drop packets in order to signal congestion (i.e., Active Queue Management algorithms)}

We can define a universal packet scheduling algorithm (hereafter UPS) in two ways, depending on our viewpoint on the problem.  From a theoretical perspective, we call a packet scheduling algorithm {\em universal} if it can replay any {\em schedule} (the set of times at which packets arrive to and exit from the network) produced by any other scheduling algorithm. This is not of practical interest, since such schedules are not typically known in advance, but it offers a theoretically rigorous definition of universality that (as we shall see) helps illuminate its fundamental limits (i.e., which scheduling algorithms have the flexibility to serve as a UPS, and why).

From a more practical perspective, we say a packet scheduling algorithm is universal if it can achieve different desired performance objectives (such as fairness, reducing tail latency, minimizing flow completion times). In particular, we require that the UPS should match the performance of the best known scheduling algorithm for a given performance objective. 

The notion of universality for packet scheduling might seem esoteric, but we think it helps clarify some basic questions. If there exists no UPS then we should \emph{expect} to design new scheduling algorithms as performance objectives evolve. Moreover, this would make a strong argument for switches being equipped with programmable packet schedulers so that such algorithms could be more easily deployed (as argued in \cite{nsb}; in fact, it was the eloquent argument in this paper that caused us to initially ask the question about universality). 

However, if there is indeed a UPS, then it changes the lens through 
which we view the design and evaluation of packet scheduling algorithms: e.g., rather than asking whether a new scheduling algorithm meets a performance objective, we should ask whether it is easier/cheaper to implement/configure than the UPS (which could also meet that performance objective). 
Taken to the extreme, one might even argue that the existence of a (practical) UPS 
greatly diminishes the need for programmable \emph{scheduling} hardware.\footnote{Note that the case for programmable hardware as made in recent work on P4 and the RMT switch~\cite{rmt,p4} remains: these systems target programmability in header parsing and in how a packet's processing pipeline is defined (i.e., how forwarding `actions' are applied to a packet). The P4 language does not currently offer primitives for scheduling and, perhaps more importantly, the RMT switch does not implement a programmable packet scheduler; we hope our results can inform the discussion on whether and how P4/RMT might be extended to support programmable scheduling.}
Thus, while the rest of the paper occasionally descends into scheduling minutae, the question we are asking has important practical (and intriguing theoretical) implications. 

This paper starts from the theoretical perspective, defining a formal model of packet scheduling and our notion of replayability in  \S\ref{sec:replay}. While we can prove that there is no UPS, we prove that Least Slack Time First (LSTF)~\cite{lstf} comes as close as any scheduling algorithm to achieving universality, and empirically (via simulation) find that LSTF can closely approximate the schedules of many packet scheduling algorithms.
We then take a more practical perspective in \S\ref{sec:objectives}, finding (via simulation) that LSTF is comparable to the state of the art in achieving various performance objectives. We discuss some related work in \S\ref{sec:related} and end with a discussion of open questions and future work in \S\ref{sec:open}.

\section{Theory: Replaying Schedules}
\label{sec:replay}


\subsection{Definitions and Overview}
\label{sec:definition}

\paragraphb{Network Model:} We consider a network of store-and-forward routers connected by links. The input load to the network is a fixed set of packets $\{p\in{P}\}$, their arrival times $i(p)$ (\ie when they reach the ingress router), and the path $path(p)$ each packet takes from its ingress to its egress router. We assume no packet drops, so all packets eventually exit.
Every router executes a nonpreemptive scheduling algorithm which need not be work-conserving or deterministic and may even involve oracles that know about future packet arrivals. Different routers in the network may use different scheduling logic. For each incoming load $\{(p,i(p),path(p))\}$, a collection of scheduling algorithms $\{A_{\alpha}\}$ (router $\alpha$ implements algorithm $A_{\alpha}$) will produce a set of packet output times $\{o(p)\}$ (the time a packet $p$ exits the network). We call the set $\{(path(p),i(p),o(p))\}$ a {\em schedule}. 

\sloppy

\paragraphb{Replaying a Schedule:} Applying a different collection of scheduling algorithms $\{A'_{\alpha}\}$ to the same set of packets $\{(p,i(p),path(p))\}$ produces a new set of output times $\{o'(p)\}$. We say that $\{A'_{\alpha}\}$ \emph{replays} $\{A_{\alpha}\}$ on this input if and only if $\forall p \in {P}$, $o'(p) \le o(p)$.\footnote{We allow the inequality because, if $o'(p) < o(p)$, one can delay the packet upon arrival at the egress node to ensure $o'(p) = o(p)$.} 

\paragraphb{Universal Packet Scheduling Algorithm:}
We say a schedule $\{(path(p),i(p),o(p))\}$ is {\em viable} if there is at least one collection of scheduling algorithms that produces that schedule. We say that a scheduling algorithm is \emph{universal} if it can replay {\em all} viable schedules. While we allowed significant generality in defining the scheduling algorithms that a UPS seeks to replay (demanding only that they be nonpreemptive), we insist that the UPS itself obey several practical constraints (although we allow it to be preemptive for theoretical analysis, but then quantitatively analyze the nonpreemptive version in \S\ref{sec:replaysims}):\footnote{The issue of preemption is somewhat complicated. Allowing the original scheduling algorithms to be preemptive allows packets to be fragmented, which then makes replay extremely difficult even in simple networks (with store-and-forward routers).  However, disallowing preemption in the candidate UPS overly limits the flexibility and would again make replay impossible even in simple networks.  Thus, we take the seemingly hypocritical but only theoretically tractable approach and disallow preemption in the original scheduling algorithms but allow preemption in the candidate UPS.  In practice, when we care only about approximately replaying schedules, the distinction is of less importance, and we simulate LSTF in the nonpreemptive form.} We impose three practical constraints on a UPS:

\fussy

\paragraphi{(1) Uniformity and Determinism:} A UPS must use the same deterministic scheduling logic at every router.

\paragraphi{(2) Limited state used in scheduling decisions:} We restrict a UPS to using only (i) packet headers, and (ii) static information about the network topology, link bandwidths, and propagation delays. It cannot rely on oracles or other external information. However, it can modify the header of a packet before forwarding it (resulting in {\em dynamic packet state} \cite{dps}).  

\paragraphi{(3) Limited state used in header initialization:}  
We assume that the header for a packet $p$ is initialized at its ingress node. The additional information available to the ingress for this initialization is limited to: (i) $o(p)$ from the original schedule and (ii) $path(p)$. Later, we extend the kinds of information the header initialization process can use, and find that this is a key determinant in whether one can find a UPS.

We make three observations about the above model. 
(i) It assumes greater capability at the edge than in the core, in keeping with common assumptions that the edge is capable of greater processing complexity, exploited by many architectural proposals\cite{csfq,sdnv2,fabric}.
(ii) When initializing a packet $p$'s header, the ingress can only use the input time, output time and the path information for $p$ itself, and must be \emph{oblivious}~\cite{oblivious} to the corresponding attributes for \emph{other} packets in the network. 
(iii) The key source of impracticality in our model is the assumption that the output times $o(p)$ are known at the ingress. However, a different interpretation of $o(p)$ suggests a practical application of replayability (and thus our results): {\em if we assign $o(p)$ as the ``desired'' output time for every packet $p$, then the existence of a UPS tells us that if these goals are viable then the UPS will be able to meet them.}

\subsection{Theoretical Results}
\label{sec:analysis}
For brevity, in this section we only summarize our key results. Interested readers can find detailed proofs in the appendix.

\paragraphb{Existence of a UPS under omniscient initialization:} Suppose we give the header-initialization process extensive information in the form of times $o(p, \alpha)$ which represent when $p$ was forwarded by router $\alpha$ in the original schedule. We can then insert an $n$-dimensional vector in the header of every packet $p$, where the $i^{th}$ element contains $o(p, \alpha_i)$, $\alpha_i$ being the $i^{th}$ hop in $path(p)$. Every time a packet arrives at a router, the router can pop the value at the head of the vector in its header and use that as its priority (earlier values of output times get higher priority). This can perfectly replay any viable schedule (proof in Appendix \ref{app:existenceproof}). This is not surprising, as having such detailed knowledge of the internal scheduling of the network is tantamount to knowing the scheduling algorithm itself. For reasons discussed previously, our definition limited the information available to the output time from the network as a whole, not from each individual router; we call this \emph{black-box} initialization.

\paragraphb{Nonexistence of a UPS under black-box initialization:}
 We can prove by counter-example (described in Appendix \ref{app:nonexistence}) that \emph{there is no UPS} under the conditions stated in \S\ref{sec:definition}. 
Given this result, we now ask \emph{how close can we get to a UPS?}

\paragraphb{Natural candidates for a near-UPS:} \changed{Simple priority scheduling \footnote{Simple priority scheduling is where the ingress assigns priority values to the packets and the routers simply schedule packets based on these static priority values.} can reproduce all viable schedules on a single router, so it would seem to be a natural candidate for a near-UPS. However, for multihop networks it may be important to make the scheduling of a packet dependent on what has happened to it earlier in its path.  For this, we consider Least Slack Time First (LSTF)~\cite{lstf}.
In LSTF, each packet $p$ carries its slack value in the packet header, which is initialized to $slack(p) = (o(p) - i(p) - t_{min}(p, src(p), dest(p)))$ at the ingress, where $src(p)$ is the ingress of $p$; $dest(p)$ is the egress of $p$; $t_{min}(p,\alpha,\beta)$ is the time $p$ takes to go from router $\alpha$ to router $\beta$ in an empty network. The slack value, therefore, indicates the maximum queueing time that the packet could tolerate without violating the replay condition. Each router, then, schedules the packet which has the least remaining slack at the time when its last bit is transmitted.} Before forwarding the packet, the router overwrites the slack value in the packet's header with its remaining slack (\ie the previous slack time minus how much time it waited in the queue before being transmitted). \footnote{There are other ways to implement this algorithm, such as using additional state in the routers and having a static packet header as in Earliest Deadline First (EDF), but here we chose to use an approach with dynamic packet state. We provide more details about EDF and prove its equivalence to LSTF in Appendix \ref{app:edf-lstf-equivalence}.}

\sloppy

\paragraphb{Key Results:} Our analysis shows that the difficulty of replay is determined by the number of \emph{congestion points}, where a \emph{congestion point} is defined as a node where a packet is forced to ``wait'' during a given schedule.
Our theorems show the following key results:

\paragraphi{1.} Priority scheduling can replay all viable schedules with no more than one congestion point per packet, and there are viable schedules with no more than two congestion points per packet that it cannot replay.
(Proof in Appendix~\ref{app:prioritiesfailure}.)

\paragraphi{2.} LSTF can replay all viable schedules with no more than two congestion points per packet, and there are viable schedules with no more than three congestion points per packet that it cannot replay.
 (Proof in Appendix~\ref{app:lstfproof}.)

\paragraphi{3.} There is no scheduling algorithm (obeying the aforementioned constraints on UPSs) that can replay {\em all} viable schedules with no more than three congestion points per packet, and the same holds for larger numbers of congestion points.
 (Proof in Appendix~\ref{app:nonexistence}.)

\paragraphb{Main Takeaway:} {\em LSTF is closer to being a UPS than simple priority scheduling, and no other candidate UPS can do better in terms of handling more congestion points.} 

\paragraphb{Intuition:} The reason why LSTF is superior to priority scheduling is clear: by carrying information about previous delays in the packet header (in the form of the {\em remaining} slack value), LSTF can ``make up for lost time'' at later congestion points, whereas for priority scheduling packets with low priority might get repeatedly delayed (and thus miss their target output times). LSTF can always handle up to two congestion points per packet because, for this case, each congestion point is either the first or the last point where the packet waits; we can prove that any extra delay seen at the first congestion point during the replay can be naturally compensated for at the second. With three or more congestion points there is no way for LSTF (or any other packet scheduler) to know how to allocate the slack among them; one can create counter-examples where unless the scheduling algorithm makes precisely the right choice in the earlier congestion points, at least one packet will miss its target output time.

\subsection{Empirical Results}
\label{sec:replaysims}

\begin{table}
\centering
\renewcommand{\arraystretch}{0.8}
\newcolumntype{P}[1]{>{\centering\arraybackslash}p{#1}}
\scriptsize
\begin{tabular}[b]{|c|P{1.2cm}|P{1.2cm}|P{0.9cm}P{0.9cm}|}
\hline
\textbf{Topology} & \textbf{Link Utilization} & \textbf{Scheduling Algorithm} &  \multicolumn{2}{P{2.0cm}|}{\textbf{Fraction of packets overdue}} \\
& & & Total & $>T$  \\
\hline 
 &  &  &  \multicolumn{2}{c|}{}  \\
I2 1Gbps-10Gbps & 70\% & Random & 0.0021 & 0.0002 \\
 &  &  &  \multicolumn{2}{c|}{} \\
 \hline
  &  &  &  \multicolumn{2}{c|}{} \\
\multirow{3}{*}{I2 1Gbps-10Gbps} & 10\% & \multirow{3}{*}{Random} & 0.0007 & 0.0  \\
& 30\% & & 0.0281 & 0.0017  \\
& 50\% & & 0.0221 & 0.0002  \\
& 90\% &  & 0.0008 & $4 \times 10^{-6}$ \\
 &  &  &  \multicolumn{2}{c|}{} \\
 \hline 
   &  &  &  \multicolumn{2}{c|}{} \\
I2 1Gbps-1Gbps & \multirow{2}{*}{70\%} & \multirow{2}{*}{Random} & 0.0204 & $8 \times 10^{-6}$  \\
I2 10Gbps-10Gbps & & & 0.0631 & 0.0448  \\
 &  &  &  \multicolumn{2}{c|}{}  \\
  \hline
     &  &  &  \multicolumn{2}{c|}{} \\
     RocketFuel & \multirow{2}{*}{70\%} & \multirow{2}{*}{Random} & 0.0246 & 0.0063  \\
Datacenter & & & 0.0164 & 0.0154  \\
 &  &  &  \multicolumn{2}{c|}{} \\
  \hline
  &  &  &  \multicolumn{2}{c|}{} \\
\multirow{5}{*}{I2 1Gbps-10Gbps} & \multirow{5}{*}{70\%} & FIFO & 0.0143 & 0.0006  \\
& & FQ & 0.0271 & 0.0002  \\
& & SJF & 0.1833 & 0.0019  \\
& & LIFO & 0.1477 & 0.0067  \\
& & FQ/FIFO+ & 0.0152 & 0.0004  \\
 &  &  &  \multicolumn{2}{c|}{}  \\
\hline
\end{tabular}
\caption{LSTF Replayability Results across various scenarios. $T$ represents the transmission time of the bottleneck link.}
\vspace{-5pt}
\label{fig:lstfreplayres}
\end{table}

The previous section clarified the theoretical limits on a {\em perfect} replay. Here we investigate, via ns-2 simulations~\cite{ns2}, how well (a nonpreemptable version of) LSTF can {\em approximately} replay schedules in realistic networks. 

\paragraphb{Experiment Setup:} \paragraphi{Default.} We use a simplified Internet-2 topology~\cite{internet2}, identical to the one used in~\cite{rc3} (consisting of 10 routers and 16 links in the core). We connect each core router to $10$ edge routers using $1$Gbps links and each edge router is attached to an end host via a $10$Gbps link.\footnote{We use higher than usual access bandwidths for our default setup to increase the stress on the schedulers in the routers. We also present results for smaller access bandwidths, which have better replay performance.} The number of hops per packet is in the range of 4 to 7, excluding the end hosts. We refer to this topology as I2:1Gbps-10Gbps. Each end host generates UDP flows using a Poisson inter-arrival model, and our default scenario runs at 70\% utilization. The flow sizes are picked from a heavy-tailed distribution~\cite{bufferbloat,adityaflowsizes}. Since our focus is on packet scheduling, not dropping policies, we use large buffer sizes that ensure no packet drops.

\paragraphi{Varying parameters.} We tested a wide range of experimental scenarios \eat{by varying the following parameters: access vs core link bandwidths, network utilization, network topology, network scale and traffic workloads. We} and present results for a small subset here: (1) the default scenario with network utilization varied from 10-90\% (2) the default scenario but with 1Gbps link between the endhosts and the edge routers (I2:1Gbps-1Gbps) and with 10Gbps links between the edge routers and the core (I2:10Gbps-10Gbps) and (3) the default scenario applied to two different topologies, a bigger Rocketfuel topology~\cite{rocketfuel} (with 83 routers and 131 links in the core) and a full bisection bandwidth datacenter fat-tree topology from~\cite{pfabric} (with 10Gbps links). Note that our other results were generally consistent with those presented here.

\paragraphi{Scheduling algorithms.} Our default case, which we expected to be hard to replay, uses completely arbitrary schedules produced by a \emph{random} scheduler (which picks the packet to be scheduled randomly from the set of queued up packets). We also present results for more traditional packet scheduling algorithms: FIFO, LIFO, fair queuing~\cite{fq}, SJF (shortest job first using priorities), and a scenario where half of the routers run FIFO+~\cite{fifoplus} and the other half run fair queuing.


\paragraphb{Evaluation Metrics:} We consider two metrics. First, we measure the fraction of packets that are overdue (i.e., which do not meet the original schedule's target). Second, to capture the \emph{extent} to which packets fail to meet their targets, we measure the fraction of packets that are overdue by more than a threshold value $T$, where $T$ is one transmission time on the bottleneck link ($\approx 12{\mu}s$ for 1Gbps). We pick this value of $T$ both because it is sufficiently small that we can assume being overdue by this small amount is of negligible practical importance, and also because this is the order of violation we should expect given that our implementation of LSTF is non-preemptive.\eat{, as opposed to the assumption of preemption in our analysis.} 

\begin{figure}[t]
		\centering
		\includegraphics[width=0.45\textwidth]{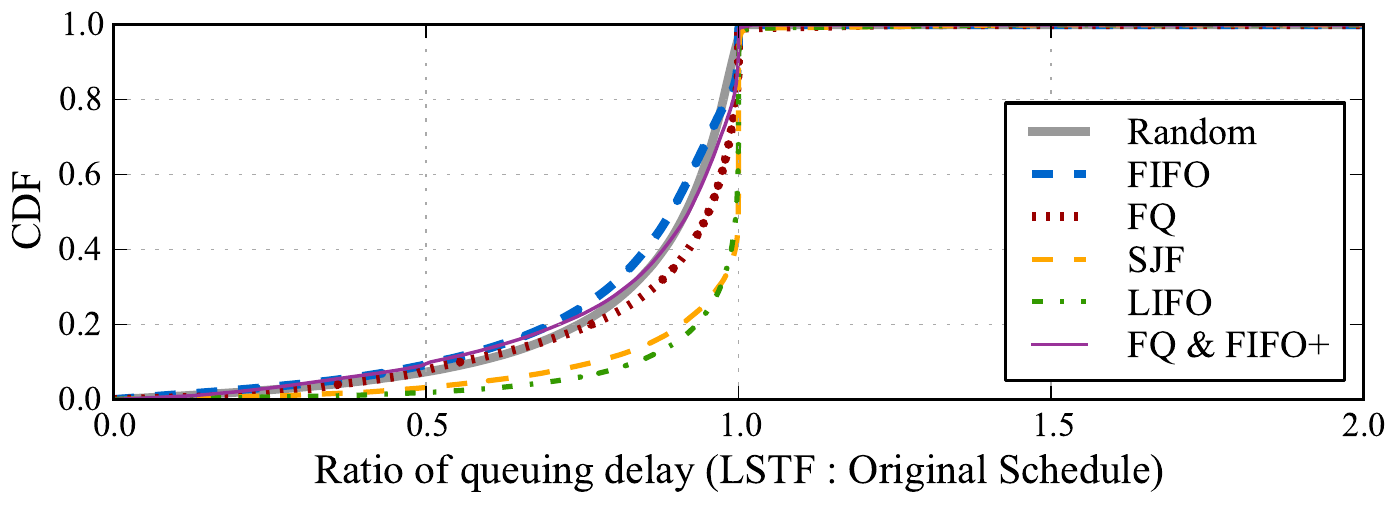}
\caption{Ratio of queuing delay with varying packet scheduling algorithms, on the default Internet-2 topology at 70\% utilization.}
\vspace{-5pt}
\label{fig:varyPS}
\end{figure}

\paragraphb{Results:} Table~\ref{fig:lstfreplayres} shows the simulation results for LSTF replay for various scenarios, which we now discuss.

\paragraphb{(1) Replayability.} Consider the column showing the fraction of packets overdue. In all but three cases (we examine these shortly) over 97\% of packets meet their target output times. In addition, the fraction of packets that did not arrive within $T$ of their target output times is much smaller; e.g., even in the worst case of SJF scheduling (where 18.33\% of packets failed to arrive by their target output times), only 0.19\% of packets are overdue by more than $T$. 
Most setups perform substantially better: e.g., in our default setup with Random scheduling, only 0.21\% of packets miss their targets and only 0.02\% are overdue by more than $T$. Hence, we conclude that even without preemption LSTF achieves good (but not perfect) replayability under a wide range of scenarios.

\paragraphb{(2) Effect of varying network utilization.} 
The second row in Table~\ref{fig:lstfreplayres} shows the effect of varying network utilization. We see that at 10\% utilization, LSTF achieves exceptionally good replayability with a total of only 0.07\% of packets overdue. Replayability deteriorates as utilization is increased to 30\% but then (surprisingly) improves again as utilization increases. This improvement occurs because with increasing utilization, the amount of queuing (and thus the average slack across packets) in the original schedule also increases, providing more room for slack re-adjustments when packets wait longer at queues seen early in their paths during the replay. We observed this trend in all our experiments though the exact location of the ``low point'' varied across settings. 

\paragraphb{(3) Effect of varying link bandwidths.} 
The third row shows the effect of changing the relative values of access {\it vs.} core links. We see that while decreasing access link bandwidth (I2:1Gbps-1Gbps) resulted in a much smaller fraction of packets being overdue by more than $T$ (0.0008\%), increasing the edge-to-core link bandwidth (I2:10Gbps-10Gbps) resulted in a significantly higher fraction (4.48\%). For I2:1Gbps-1Gbps, packets are paced by the endhost link, resulting in few congestion points thus improving LSTF's replayability. In contrast, with I2:10Gbps-10Gbps, both the access and edge links have a higher bandwidth than most core links; hence \eat{ not only does the number of congestion points a packet may encounter increases, but also }packets (that are no longer paced at the endhosts or the edges) arrive at the core routers very close to one another and the effect of one packet being overdue \emph{cascades} over to the following packets.  

\paragraphb{(4) Effect of varying topology.}
The fourth row in Table~\ref{fig:lstfreplayres} shows our results using different topologies. LSTF performs well in both cases: only 2.46\% (Rocketfuel) and 1.64\% (datacenter) of packets fail replay. These numbers are still somewhat higher than our default case. The reason for this is similar to that for the I2:10Gbps-10Gbps topology -- all links in the datacenter topology are set to 10Gbps, while half of the core links in the Rocketfuel topology are set to have bandwidths smaller than the access links.


\eat{ 
 10Gbps, due to the same reason. 

In light of the above results, we elaborate on how link bandwidths are set in each. In the Rocketfuel topology less than half of the core links had capacities smaller than the access link capacity,

has only 0.63\% packets are overdue by a duration more than $T$. This shows that the ratio between the capacities of the access and the core links (which determines the \emph{burstiness} of the traffic), is a bigger performance factor than the size of the topology.\fixme{I need to reread}
}

\paragraphb{(5) Varying Scheduling Algorithms.}
Row five in Table~\ref{fig:lstfreplayres} shows LSTF's replay results for different scheduling algorithms. We see that LSTF performs well for FIFO, FQ, and even the combination of FIFO$+$ and FQ; with fewer than 0.06\% of packets being overdue by more than $T$. 
SJF and LIFO fare worse with 18.33\% and 14.77\% of packets failing replay (although only 0.19\% and 0.67\% of packets are overdue by more than $T$ respectively). The reason stems from two factors: (1) for these algorithms a larger fraction of packets have a very small slack value (as one might expect from the scheduling logic which produces a larger skew in the slack distribution), and (2) for these packets with small slack values, LSTF \emph{without preemption} is often unable to ``compensate'' for misspent slack that occurred earlier in the path. To verify this intuition, we extended our simulator to support preemption and repeated our experiments: with preemption, the fraction of packets that failed replay dropped to 0.24\% (from 18.33\%) for SJF and to 0.25\% (from 14.77\%) for LIFO.



\paragraphb{(6) End-to-end (Queuing) Delay.}
Our results so far evaluated LSTF in terms of measures that we introduced to test universality. We now evaluate LSTF using the more traditional metric of packet delay, focusing on the queueing delay a packet experiences. Figure~\ref{fig:varyPS} shows the CDF of the ratios of the queuing delay that a packet sees with LSTF to the queuing delay that it sees in the original schedule, for varying packet scheduling algorithms. We were surprised to see that most of the packets actually have a smaller queuing delay in the LSTF replay than in the original schedule. This is because LSTF eliminates ``wasted waiting'', in that it never makes packet A wait behind packet B if packet B is going to have significantly more waiting later in its path.

\eat{ 
For other schedulers too, preemption with LSTF reduced the total fraction of packets that were overdue, though the fraction of packets that were overdue by more than $T$ saw very little change.  \fixme{fix this paragraph}
}

\paragraphb{(7) Comparison with Priorities.} 
\changed{To provide a point of comparison, we did a replay using simple priorities for our default scenario, where the priority for a packet $p$ is set to $o(p)$ (which seemed most intuitive to us). As expected, the resulting replay performance is much worse than that with LSTF: 21\% packets are overdue in total (vs 0.21\% with LSTF), with 20.69\% being overdue by more than $T$ (vs 0.02\% with LSTF).}


\paragraphb{Summary:} We observe that, in almost all cases, less than 1\% of the packets are overdue with LSTF by more than $T$. The replay performance initially degrades and then starts improving as the network utilization increases. The distribution of link speeds has a bigger influence on the replay results than the scale of the topology. Replay performance is better for scheduling algorithms that produce a smaller skew in the slack distribution. \changed{LSTF replay performance is significantly better than simple priorities replay performance, with the most intuitive priority assignment.}

\section{Practical: Achieving Various Objectives}
\label{sec:objectives}

In this section we look at how LSTF can be used \emph{in practice} to meet three popular network-wide objectives: minimizing mean flow completion time, minimizing tail packet delays, and fairness. Instead of using the knowledge of a given previous schedule (as done in \S\ref{sec:replaysims}), we now use certain heuristics (described below) to assign the slacks.

For each objective, we first describe the slack initialization heuristic and then present some ns-2 simulation results on how LSTF performs relative to the state-of-the-art scheduling algorithm on the I2 1Gbps-10Gbps topology running at 70\% average utilization.\footnote{We have run our simulations in a wide variety of scenarios and find similar results to what we present here.} The switches have finite buffers (packets with the highest slack are dropped when the buffer is full).


\subsection{Mean Flow Completion Time}

\begin{figure}[t]
\centering
\includegraphics[width=0.45\textwidth]{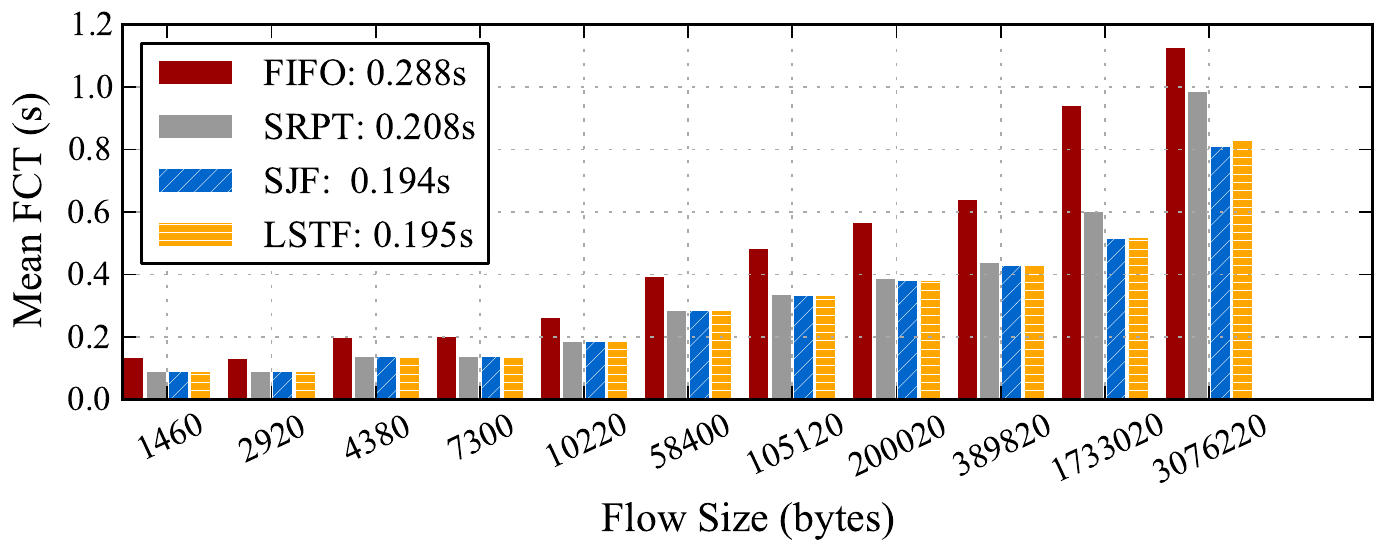}
\caption{Mean FCT bucketed by flow size for the Internet2 topology with 70\% utilization. The legend indicates the mean FCT across all flows. }
\vspace{-5pt}
\label{fig:avgFCT}
\end{figure}

While there have been several proposals on how to minimize flow completion time (FCT) via the transport protocol~\cite{whyfct, rc3}, here we focus on scheduling's impact on FCT. In \cite{pfabric} it is shown that (i) Shortest Remaining Processing Time (SRPT) is close to optimal for minimizing the mean FCT and (ii) Shortest Job First (SJF) produces results similar to SRPT for realistic heavy-tailed distribution. Thus, these are the two algorithms we use as benchmarks.

\paragraphb{Slack Initialization:} 
The slack for a packet $p$ is initialized as $slack(p) = fs(p) * D$, where $fs(p)$ is the size of the flow to which the packet $p$ belongs and $D$ is a value much larger than the delay seen by any packet in the network ($D = 1$ sec in our simulations). 

\paragraphb{Evaluation:} 
We use TCP flows with a 5MB buffer in each router (which is equal to the average delay-bandwidth product for the Internet2 topology we are using). Figure~\ref{fig:avgFCT} compares LSTF with FIFO, SJF and SRPT with \emph{starvation prevention} as in \cite{pfabric} \footnote{The router always schedules the earliest arriving packet of the flow which contains the highest priority packet.}. 
 SJF has a slightly better performance than SRPT, both resulting in a significantly lower mean FCT than FIFO. LSTF's performance is nearly the same as SJF.

\subsection{Tail Packet Delays}

\begin{figure}[t]

\centering
\includegraphics[width=0.45\textwidth]{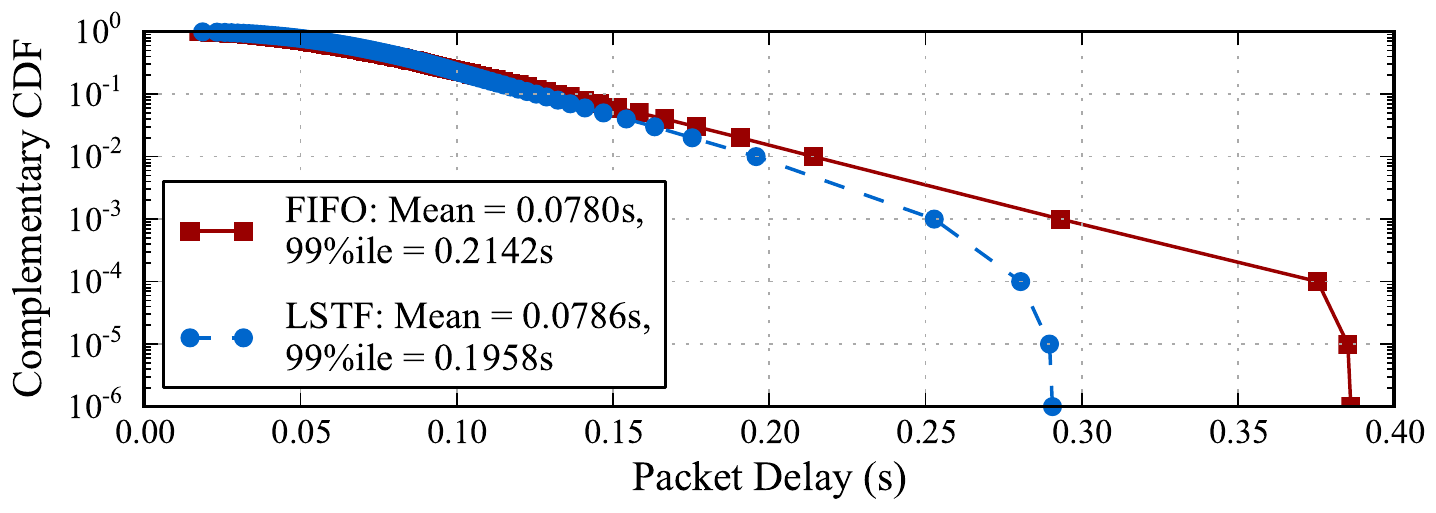}

\caption{Tail packet delays for LSTF compared to FIFO. The mean and 99\%ile packet delay values are indicated in the legend.}
\vspace{-5pt}
\label{fig:tailDelay}
\end{figure}

Clark et. al. \cite{fifoplus} proposed the FIFO+ algorithm for minimizing the tail packet delays in multi-hop networks, where packets are prioritized at a router based on the amount of queuing delay they have seen at their previous hops.

\paragraphb{Slack Initialization:} All incoming packets are initialized with the same slack value (1 sec in our simulations). This makes LSTF identical to FIFO+. 

\paragraphb{Evaluation:} We compare our LSTF policy (which, with the above slack initialization, is identical to FIFO+) with FIFO. 
We present our results using UDP flows, which ensures that the input load remains the same in both cases, allowing a fair comparison for the in-network packet-level behaviour across the two scheduling policies. Figure~\ref{fig:tailDelay} shows our results. With LSTF, packets that have traversed through more number of hops, and have therefore spent more slack in the network, get preference over shorter-RTT packets that have traversed through fewer hops. While this might produce a slight increase in the mean packet delay, it reduces the tail. This is in-line with the observations made in~\cite{fifoplus}.

\subsection{Fairness}

Fairness is the most challenging objective to achieve with LSTF, but we show that it can achieve \emph{asymptotic} fairness (i.e. eventual convergence to the fair-share rate).

\paragraphb{Slack Initialization:} Our approach is inspired from~\cite{virtualClocks}. We assign $slack = 0$ to the first packet of the flow and the slack of any subsequent packet $p_i$ is then initialized as:
\begin{align*}
slack(p_i) = max\Big(0, slack(p_{i-1}) + \frac{1}{r_{est}} - \big(i(p_i) - i(p_{i-1})\big)\Big) 
\end{align*}
where $i(p)$ is the arrival time of the packet $p$ at the ingress and $r_{est}$ is an estimate of the fair-share rate $r^*$. 
We show that the above heuristic leads to asymptotic fairness, for {\em any} value of $r_{est}$ that is less than $r^*$, as long as all flows use the same value. A reasonable value of $r_{est}$ can be estimated using knowledge about the network topology and traffic matrices, though we leave a detailed exploration of this to future work. We can also extend the slack assignment heuristic to achieve weighted fairness by using different values of $r_{est}$ for different flows, in proportion to the desired weights.

\begin{figure}[t] 
\centering
\includegraphics[width=0.45\textwidth]{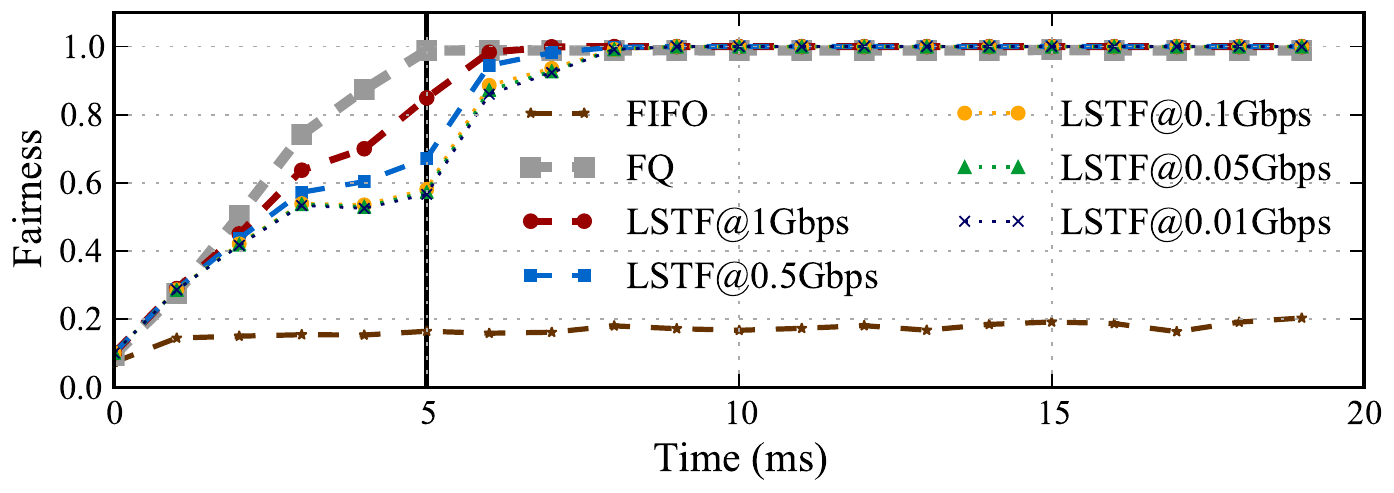}
\caption{Fairness for long-lived flows on Internet2 topology. The legend indicates the value of $r_{est}$ used for LSTF slack initialization.} 
\vspace{-5pt}
\label{fig:macroFairness} 
\end{figure}

\paragraphb{Evaluation:}
We evaluate the asymptotic fairness property by running our simulation on the Internet2 topology with 10Gbps edges, such that all the congestion is happening at the core. However, we reduce the propagation delay, to make the experiment more scalable, while the buffer size is kept large so that the fairness is dominated by the scheduling policy. We start 90 long-lived TCP flows with a random jitter in the start times ranging from 0-5ms. The topology is such that the fair share rate of each flow on each link in the core network (which is shared by up to 13 flows) is around 1Gbps. We use different values for $r_{est} \le 1$Gbps for computing the initial slacks and compare our results with fair queuing (FQ). Figure~\ref{fig:macroFairness} shows the fairness computed using Jain's Fairness Index~\cite{jain}, from the throughput each flow receives per millisecond. Since we use the throughput received by each of the 90 flows to compute the fairness index, it reaches 1 with FQ only at 5ms, after all the flows have started. We see that LSTF is able to converge to perfect fairness, even when $r_{est}$ is 100X smaller than $r^*$. It converges slightly sooner when $r_{est}$ is closer to $r^*$, though the subsequent differences in the time to convergence decrease with decreasing values of $r_{est}$.

\section{Related Work}
\label{sec:related}


The real-time scheduling literature has studied optimality of scheduling algorithms\footnote{where a scheduling algorithm is said to be optimal if it can (feasibly) schedule a set of tasks that can be scheduled by any other scheduling algorithm.} (in particular EDF and LSTF) for single and multiple processors~\cite{edf, lstf}. Liu and Layland~\cite{edf} proved the optimality of EDF for a single processor in hard real-time systems. LSTF was then shown to be optimal for single-processor scheduling as well, while being more effective than EDF (though not optimal) for multi-processor scheduling~\cite{lstf}. In the context of networking,~\cite{pifo} provides theoretical results on emulating the schedules produced by a single output-queued switch using a combined input-output queued switch with a smaller speed-up of at most two. To the best of our knowledge, the optimality or universality of a scheduling algorithm for a network of inter-connected resources (in our case, switches) has never been studied before.

\eat{The scheduling algorithms targeted for this purpose are restricted to those that can be expressed using the ``push-in-first-out'' (PIFO) primitive, where a packet is inserted in an arbitrary location in the queue and the packet at the head of the queue is scheduled. This class includes scheduling algorithms such as strict priorities, weighted fair queuing and even LSTF.}

\fixme{Ate the part about use of EDF for meeting deadlines and LSTF for FIFO+.}

A recent paper \cite{nsb} proposed programmable hardware in the dataplane for packet scheduling and queue management, in order to achieve various network objectives without the need for physically replacing the hardware. It uses simulation of three schemes (FQ, CoDel$+$FQ, CoDel$+$FIFO) competing on three different metrics to show that there is no ``silver bullet'' solution. 
As mentioned earlier, our work is inspired by the questions the authors raise; we adopt a broader view of scheduling in which packets can carry dynamic state leading to the results presented here. 

\section{Open Questions and Future Work} 
\label{sec:open}

\paragraphb{Theoretical Analysis:}
Our work leaves several theoretical questions unanswered, including the following.  First, we showed existence of a UPS with omniscient header initialization, and nonexistence with limited-information initialization. {\em What is the least information we can use in header initialization in order to achieve universality?}
Second, we showed that, the fraction of overdue packets is small, and most are only overdue by a small amount during an LSTF replay. {\em Are there tractable bounds on both the number of overdue packets and/or their degree of lateness?} \changed{Finally, while we have a formal analysis of LSTF's ability to replay a given schedule, we do not yet have any formal model for the scope of LSTF in meeting various objectives in practice. \emph{Can one describe the class of performance objectives that LSTF can meet?}}

\eat{\paragraphb{Measuring the Number of Congestion Points in the Internet:} As we noted in \S\ref{sec:analysis}, having more than two congestion points per packet in the network for a given input load is a necessary condition for a replay failure with LSTF. There has been previous work~\cite{measurecp} on measuring the number of occurrences of more than one congestion point between two nodes in the Internet using packet pairs. As a part of our future work, we would like to extend this methodology to measure the occurrences of more than two congestion points.
\fixme{Should we remove this, to get within page limit?}}


\paragraphb{Real Implementation:} 
\eat{Hardware implementation of LSTF at a router would comprise of two basic components - scheduling the least slack packet and updating the slack of the packet being scheduled. LSTF scheduling has a nice property that the difference in the slack value of any two packets waiting at a queue remains the same throughout, which means that the relative ordering between them also remains the same.} 
\changed{We need to show the feasibility of implementing LSTF in hardware. However, we can prove that LSTF execution \emph{at a particular router} is no more complex than the execution of fine-grained priorities, which can be carried out in almost constant time using specialized data-structures such as pipelined heap (p-heap)~\cite{pheap1, pheap2}.}


\eat{Another concern with respect to real implementation is how to encode the slack values in the packets, which involves a space vs accuracy trade-off.} 

\fixme{Ate the part about ``Achieving other Network Objectives}

\paragraphb{Incorporating Feedback:}
Typically congestion control involves endhosts reacting to network feedback, which can be implicit (e.g., packet drops by Active Queue Management schemes~\cite{red,codel}) or explicit (e.g., ECN markings~\cite{ecn} or rate allocations schemes such as RCP~\cite{whyfct} and XCP~\cite{xcp}). It is unclear whether it is necessary to incorporate such feedback mechanisms in our notion of universality, and if so how. \eat{It is possible that one can achieveOne possibility is to rely on an end-to-end delay-based congestion control to achieve the power objective without using any switch support. Proposals for delay-based congestion control have existed in the context of wide-area networks for a long time~\cite{vegas,fasttcp} and more recently, it was shown to be feasible (and in fact better than ECN) for datacenters as well~\cite{dx, timely}. An alternative option is to run LSTF in conjunction with a feedback based scheme at the routers. However, we defer the exploration of these alternatives to future work.}   

\section{Acknowledgements}

\changed{We are thankful to Prof. Satish Rao for his very helpful tips regarding the theoretical aspects of this work. We would also like to thank Prof. Ion Stoica, Kay Ousterhout, Justine Sherry, Aurojit Panda and our anonymous HotNets reviewers for their thoughtful feedback. This work was in part made possible by generous financial support from Intel Research.}

\bibliographystyle{abbrv} 
\bibliography{main}

\begin{thebibliography}{10}

\bibitem{internet2}
{I}nternet2.
\newblock \url{http://www.internet2.edu/}.

\bibitem{ns2}
{The Network Simulator NS-2}.
\newblock \url{http://www.isi.edu/nsnam/ns/}.

\bibitem{pfabric}
M.~Alizadeh, S.~Yang, M.~Sharif, S.~Katti, N.~McKeown, B.~Prabhakar, and
  S.~Shenker.
\newblock {pFabric: Minimal Near-optimal Datacenter Transport}.
\newblock In {\em Proc. ACM SIGCOMM}, 2013.

\bibitem{bufferbloat}
M.~Allman.
\newblock {Comments on bufferbloat}.
\newblock {\em ACM SIGCOMM Computer Communication Review}, 2013.

\bibitem{adityaflowsizes}
T.~Benson, A.~Akella, and D.~Maltz.
\newblock {Network Traffic Characteristics of Data Centers in the Wild}.
\newblock In {\em Proc. ACM IMC}, 2012.

\bibitem{pheap1}
R.~Bhagwan and B.~Lin.
\newblock {Fast and Scalable Priority Queue Architecture for High-Speed Network
  Switches}.
\newblock In {\em Proc. IEEE Infocom}, 2000.

\bibitem{p4}
P.~Bosshart, D.~Daly, G.~Gibb, M.~Izzard, N.~McKeown, J.~Rexford,
  C.~Schlesinger, D.~Talayco, A.~Vahdat, G.~Varghese, and D.~Walker.
\newblock {P4: Programming Protocol-independent Packet Processors}.
\newblock {\em ACM SIGCOMM Computer Communication Review}, 2014.

\bibitem{rmt}
P.~Bosshart, G.~Gibb, H.-S. Kim, G.~Varghese, N.~McKeown, M.~Izzard, F.~Mujica,
  and M.~Horowitz.
\newblock {Forwarding Metamorphosis: Fast Programmable Match-action Processing
  in Hardware for SDN}.
\newblock In {\em Proc. ACM SIGCOMM}, 2013.

\bibitem{fabric}
M.~Casado, T.~Koponen, S.~Shenker, and A.~Tootoonchian.
\newblock {Fabric: A Retrospective on Evolving SDN}.
\newblock In {\em Proc. ACM HotSDN}, 2012.

\bibitem{pifo}
S.-T. Chuang, A.~Goel, N.~McKeown, and B.~Prabhakar.
\newblock {Matching output queueing with a combined input/output-queued
  switch}.
\newblock {\em IEEE Journal on Selected Areas in Communications}, 1999.

\bibitem{fifoplus}
D.~D. Clark, S.~Shenker, and L.~Zhang.
\newblock {Supporting Real-time Applications in an Integrated Services Packet
  Network: Architecture and Mechanism}.
\newblock {\em ACM SIGCOMM Computer Communication Review}, 1992.

\bibitem{fq}
A.~Demers, S.~Keshav, and S.~Shenker.
\newblock {Analysis and Simulation of a Fair Queueing Algorithm}.
\newblock {\em ACM SIGCOMM Computer Communication Review}, 1989.

\bibitem{whyfct}
N.~Dukkipati and N.~McKeown.
\newblock {Why Flow-Completion Time is the Right Metric for Congestion
  Control}.
\newblock {\em ACM SIGCOMM Computer Communication Review}, 2006.

\bibitem{red}
S.~Floyd and V.~Jacobson.
\newblock {Random Early Detection Gateways for Congestion Avoidance}.
\newblock {\em IEEE/ACM Trans. Netw.}, 1993.

\bibitem{oblivious}
A.~Gupta, M.~T. Hajiaghayi, and H.~R\"{a}cke.
\newblock {Oblivious Network Design}.
\newblock In {\em Proc. ACM-SIAM Symposium on Discrete Algorithm (SODA)}, 2006.

\bibitem{pheap2}
A.~Ioannou and M.~G.~H. Katevenis.
\newblock {Pipelined Heap (Priority Queue) Management for Advanced Scheduling
  in High-speed Networks}.
\newblock {\em IEEE/ACM Trans. Netw.}, 2007.

\bibitem{jain}
R.~Jain, D.-M. Chiu, and W.~Hawe.
\newblock {A Quantitative Measure Of Fairness And Discrimination For Resource
  Allocation In Shared Computer Systems}.
\newblock {\em CoRR}, 1998.

\bibitem{xcp}
D.~Katabi, M.~Handley, and C.~Rohrs.
\newblock {Congestion Control for High Bandwidth-Delay Product Networks}.
\newblock In {\em Proc. ACM SIGCOMM}, 2002.

\bibitem{lstf}
J.~Y.-T. Leung.
\newblock {A new algorithm for scheduling periodic, real-time tasks}.
\newblock {\em Algorithmica, Springer-Verlag New York Inc.}, 1989.

\bibitem{edf}
C.~L. Liu and J.~W. Layland.
\newblock {Scheduling Algorithms for Multiprogramming in a Hard-Real-Time
  Environment}.
\newblock {\em Journal of the ACM (JACM)}, 1973.

\bibitem{rc3}
R.~Mittal, J.~Sherry, S.~Ratnasamy, and S.~Shenker.
\newblock {Recursively Cautious Congestion Control}.
\newblock In {\em Proc. USENIX NSDI}, 2014.

\bibitem{codel}
K.~Nichols and V.~Jacobson.
\newblock {Controlling Queue Delay}.
\newblock {\em ACM Queue}, 2012.

\bibitem{gps}
A.~K. Parekh and R.~G. Gallager.
\newblock {A Generalized Processor Sharing Approach to Flow Control in
  Integrated Services Networks: The Single-node Case}.
\newblock {\em IEEE/ACM Trans. Netw.}, 1993.

\bibitem{sdnv2}
B.~Raghavan, M.~Casado, T.~Koponen, S.~Ratnasamy, A.~Ghodsi, and S.~Shenker.
\newblock {Software-defined Internet Architecture: Decoupling Architecture from
  Infrastructure}.
\newblock In {\em Proc. ACM HotNets}, 2012.

\bibitem{ecn}
K.~Ramakrishnan, S.~Floyd, and D.~Black.
\newblock {The Addition of Explicit Congestion Notification (ECN) to IP}.
\newblock {RFC 3168}, 2001.

\bibitem{diffserv}
{S. Blake and D. Black and M. Carlson and E. Davies and Z. Wang and W. Weiss}.
\newblock {An Architecture for Differentiated Services}.
\newblock {RFC 2475}, 1998.

\bibitem{drr}
M.~Shreedhar and G.~Varghese.
\newblock {Efficient Fair Queueing Using Deficit Round Robin}.
\newblock {\em ACM SIGCOMM Computer Communication Review}, 1995.

\bibitem{nsb}
A.~Sivaraman, K.~Winstein, S.~Subramanian, and H.~Balakrishnan.
\newblock {No Silver Bullet: Extending SDN to the Data Plane}.
\newblock In {\em Proc. ACM HotNets}, 2013.

\bibitem{rocketfuel}
N.~Spring, R.~Mahajan, and D.~Wetherall.
\newblock {Measuring ISP Topologies with Rocketfuel}.
\newblock In {\em Proc. ACM SIGCOMM}, 2002.

\bibitem{csfq}
I.~Stoica, S.~Shenker, and H.~Zhang.
\newblock {Core-stateless Fair Queueing: Achieving Approximately Fair Bandwidth
  Allocations in High Speed Networks}.
\newblock In {\em Proc. ACM SIGCOMM}, 1998.

\bibitem{dps}
I.~Stoica and H.~Zhang.
\newblock {Providing Guaranteed Services Without Per Flow Management}.
\newblock In {\em Proc. ACM SIGCOMM}, 1999.

\bibitem{virtualClocks}
L.~Zhang.
\newblock {Virtual Clock: A New Traffic Control Algorithm for Packet Switching
  Networks}.
\newblock {\em ACM SIGCOMM Computer Communication Review}, 1990.

\end{thebibliography}
\label{last-page}

\section*{Appendix}
\begin{appendices}

This section contains proofs for our theoretical results presented in \S\ref{sec:replay}. 
\section{Notations}

We use the following notations for our proofs, some of which have been already defined in the main text:

\paragraphb{Relevant nodes:}

\begin{itemize}[nolistsep,leftmargin=*]
  \item $src(p)$: Ingress of a packet $p$.
  \item $dest(p)$: Egress of a packet $p$.
\end{itemize}

\paragraphb{Relevant time notations:}
\begin{itemize}[nolistsep,leftmargin=*]
  \item $T(p, \alpha)$: Transmission time of a packet $p$ at node $\alpha$.
  \item $o(p, \alpha)$: Time when the first bit of $p$ is scheduled by node $\alpha$ in the original schedule.
  \item $o(p) = o(p,dest(p)) + T(p,dest(p))$: Time when the last bit of $p$ exits the network in the original schedule (which is non-preemptive).
  \item $o'(p)$: Time when the last bit of $p$ exits the network in the replay (which may be preemptive in our theoretical arguments).
  \item $i(p, \alpha)$ and $i'(p, \alpha)$: Time when $p$ arrives at node $\alpha$ in the original schedule and in the replay respectively.
  \item $i(p) = i(p,src(p)) = i'(p)$: Arrival time of $p$ at its ingress. This remains the same for both the original schedule and the replay. 
  \item $t_{min}(p,\alpha,\beta)$: Minimum time $p$ takes to start from node $\alpha$ and exit from node $\beta$ in an uncongested network. It therefore includes the propagation delays and the store-and-forward delays of all links in the path from $\alpha$ to $\beta$ and the transmission delays at $\alpha$ and $\beta$. Handling the edge case: $t_{min}(p,\alpha,\alpha) = T(p,\alpha)$
  \item $slack(p) = o(p) - i(p) - t_{min}(p,src(p),dest(p))$: Total slack of $p$ that gets assigned at its ingress. It denotes the amount of time $p$ can wait in the network without any of its bits getting serviced.
  \item $slack(p,\alpha,t) =  o(p) - t - t_{min}(p,\alpha,dest(p)) + T(p, \alpha)$: Remaining slack of the last bit of $p$ at time $t$ when it is at node $\alpha$. We derive this expression in Appendix~\ref{app:slackeqn}.
\end{itemize}

\paragraphb{Other miscellaneous notations}
\begin{itemize}[nolistsep,leftmargin=*]
\item $path(p,\alpha,\beta)$: The ordered set of nodes and links in the path taken by $p$ to go from $\alpha$ to $\beta$. The set also includes $\alpha$ and $\beta$ as the first and the last nodes. 
\item $path(p) = path(p, src(p), dest(p))$
\item $pass(\alpha)$: Set of packets that pass through node $\alpha$.
\end{itemize}

\section{Existence of a UPS under Omniscient Header Initialization}
\label{app:existenceproof}

\paragraph{Algorithm:} At the ingress, insert an $n$-dimensional vector in the packet header, where the $i^{th}$ element contains $o(p, \alpha_i)$, $\alpha_i$ being the $i^{th}$ hop in $path(p)$. Every time a packet $p$ arrives at the router, the router pops the value at the head of the vector in $p$'s header and uses that as the priority for $p$ (earlier values of output times get higher priority). This can perfectly replay any schedule.

\paragraph{Proof:}
We can prove that the above algorithm will result in no overdue packets (which do not meet their original schedule's target) using the following two theorems:

\paragraphb{Theorem 1:}
If for any node $\alpha$, $\exists p' \in pass(\alpha)$, such that using the above algorithm, the last bit of $p'$ exits $\alpha$ at time $(t' > (o(p',\alpha) + T(p',\alpha)))$, then $(\exists p \in pass(\alpha) \mid i'(p, \alpha) \le t' \text{ and } i'(p, \alpha) > o(p, \alpha))$.

\paragraphi{Proof by contradiction:} Consider the first such $p^* \in pass(\alpha)$ that gets late at $\alpha$ (i.e. its last bit exits $\alpha$ at time $t^* > (o(p^*,\alpha) + T(p^*,\alpha))$). Suppose the above condition is not true i.e. $(\forall p \in pass(\alpha) \mid  i'(p, \alpha) \le o(p, \alpha) \text{ or } i'(p, \alpha) > t^*)$. In other words, if $p$ arrives at or before time $t^*$, it also arrives at or before time $o(p, \alpha)$. Given that all bits of $p^*$ arrive at or before time $t^*$, they also arrive at or before time $o(p^*, \alpha)$. The only reason why the last bit of $p^*$ would wait until time $(t^* >  o(p^*,\alpha) + T(p^*,\alpha))$ in our work-conserving replay is if some other bits (belonging to higher priority packets) were being scheduled after time $o(p^*, \alpha)$, resulting in $p^*$ not being able to complete its transmission by time $(o(p^*,\alpha) + T(p^*, \alpha))$. However, as per our algorithm, any packet $p_{high}$ having a higher priority than $p^*$ at $\alpha$ must have been scheduled before $p^*$ in the original schedule, implying that $(o(p_{high}, \alpha) + T(p_{high}, \alpha)) \le o(p^*, \alpha)$. \footnote{Given that the original schedule is non-preemptible, the next packet gets scheduled only after the previous one has completed its transmission.} Therefore, some bits of $p_{high}$ being scheduled after time $o(p^*,\alpha)$, implies them being scheduled after time $(o(p_{high}, \alpha) + T(p_{high}, \alpha))$. This means that $p_{high}$ is already late and contradicts our assumption that $p^*$ is the first packet to get late. Hence proved that if for any node $\alpha$, $\exists p' \in pass(\alpha)$, such that using the above algorithm, the last bit of $p'$ exits $\alpha$ at time $(t' > (o(p',\alpha) + T(p',\alpha)))$, then $(\exists p \in pass(\alpha) \mid i'(p, \alpha) \le t' \text{ and } i'(p, \alpha) > o(p, \alpha))$

\paragraphb{Theorem 2:} $\forall \alpha, (\forall p \in pass(\alpha) \mid i'(p,\alpha) \le i(p,\alpha))$.

\paragraphi{Proof by contradiction:} Consider the first time when some packet $p^*$ arrives late at some node $\alpha^*$ (i.e. $i'(p^*,\alpha^*) > i(p^*,\alpha^*)$). In other words, $\alpha^*$ is the first node in the network to see a late packet arrival, and $p^*$ is the first late arriving packet. Let $\alpha_{prev}$ be the node visited by $p^*$ just before arriving at $\alpha^*$. $p^*$ can arrive at a time later than $i(p^*, \alpha^*)$ at $\alpha^*$ only if the last bit of $p^*$ exits $\alpha_{prev}$ at time $t_{prev} > o(p^*, \alpha_{prev}) + T(p^*, \alpha_{prev})$. As per Theorem 1 above, this is possible only if some packet $p'$ (which may or may not be same as $p^*$) arrives at $\alpha_{prev}$ at time $i'(p', \alpha_{prev}) > o(p',\alpha_{prev}) \ge i(p',\alpha_{prev})$ and $i'(p', \alpha_{prev}) \le t_{prev} < i'(p^*, \alpha^*)$. This contradicts our assumption that $\alpha^*$ is the first node to see a late arriving packet. Therefore, $\forall \alpha, (\forall p \in pass(\alpha) \mid i'(p,\alpha) \le i(p,\alpha))$.

Combining the two theorems above: Since $\forall \alpha (\forall p \in pass(\alpha) \mid i'(p,\alpha) \le i(p,\alpha))$, with the above algorithm, $\forall \alpha(\forall p \in pass(\alpha))$, all bits of $p$ exit $\alpha$ before $(o(p,\alpha) + T(p,\alpha))$. Therefore, the algorithm can perfectly replay any viable schedule.

\section{Nonexistence of a UPS under black-box initialization}
\label{app:nonexistence}

\begin{figure}[t]
\centering

\begin{subfigure}
\centering
\includegraphics[width=0.45\textwidth]{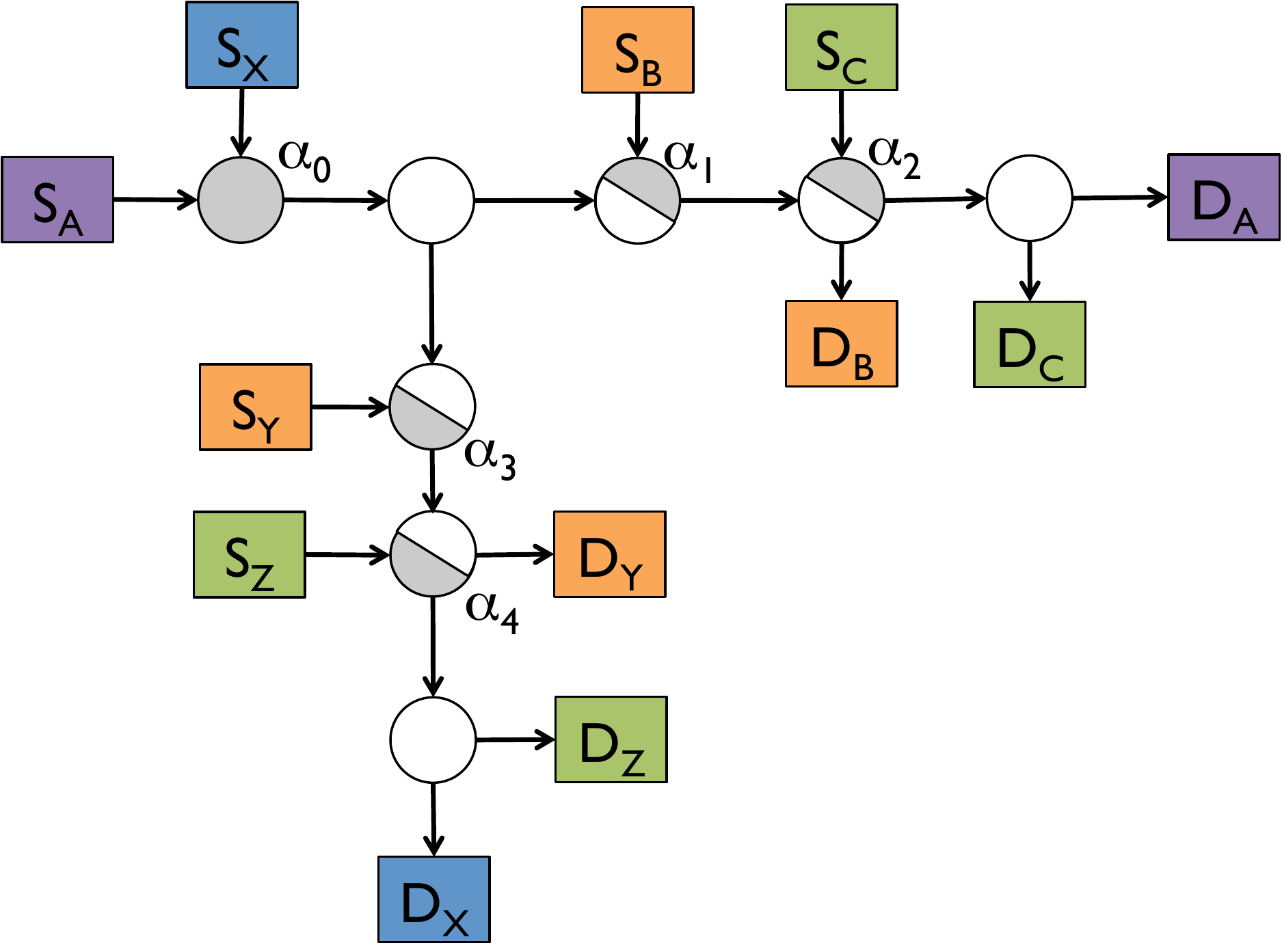}
\end{subfigure}

\small
\begin{tabular}{|cc|}
\hline
{\bf Node} & {\bf Packet(arrival time, scheduling time)} \\
\hline
\multicolumn{2}{|c|}{\emph{Case 1}} \\
\hline
$\alpha_0$ & $a(\textcolor{blue}{\pmb{0}},0)$; $x(\textcolor{blue}{\pmb{0}},1)$ \\
$\alpha_1$ & $a(1,1)$, $b_1(2,2)$, $b_2(3,3), b_3(4,4)$  \\
$\alpha_2$ & $c_1(2,2)$, $c_2(3,3)$; $a(2,\textcolor{blue}{\pmb{4}})$  \\
$\alpha_3$ & $x(2,2)$, $y_1(2,3)$, $y_2(3,4)$  \\
$\alpha_4$ & $z(2,2)$, $x(3,\textcolor{blue}{\pmb{3}})$  \\
\hline
\multicolumn{2}{|c|}{\emph{Case 2}} \\
\hline
$\alpha_0$ &  $x(\textcolor{blue}{\pmb{0}},0)$; $a(\textcolor{blue}{\pmb{0}},1)$ \\
$\alpha_1$ &  $a(2,2)$, $b_1(2,3)$, $b_2(3,4), b_3(4,5)$ \\
$\alpha_2$ &  $c_1(2,2)$, $c_2(3,3)$, $a(3,\textcolor{blue}{\pmb{4}})$ \\
$\alpha_3$ &  $x(1,1)$, $y_1(2,2)$, $y_2(3,3)$ \\
$\alpha_4$ &  $z(2,2)$, $x(2,\textcolor{blue}{\pmb{3}})$ \\
\hline
\end{tabular}

\caption{Example showing non-existence of a UPS with Blackbox Initialization. A packet represented by $p$ belongs to flow $P$, with ingress $S_P$ and egress $D_P$, where $P \in \{A, B, C, X, Y, Z\}$.  For simplicity assume all packets are of the same size and all links have a propagation delay of zero. All uncongested routers (white), ingresses and egresses have a transmission time of zero. The congestion points (shaded grey) have transmission times of $T = 1$ unit.}
\label{fig:no-univ-algo}
\end{figure}

\paragraphb{Proof by counter-example:} Consider the example shown in Figure~\ref{fig:no-univ-algo}. For simplicity, assume all the propagation delays are zero, the transmission time for each congestion point (shaded in grey) is 1 unit and the uncongested (white) routers have zero transmission time. \footnote{These assignments are made for simplicity of understanding. The example will hold for any reasonable value of propagation and transmission delays.} All packets are of the same size.

The table illustrates two cases. For each case, a packet's arrival and scheduling time (the time when the packet is scheduled by the router) at each node through which it passes are listed.  A packet represented by $p$ belongs to flow $P$, with ingress $S_P$ and egress $D_P$, where $P \in \{A, B, C, X, Y, Z\}$. The packets have the same $path$ in both cases. For example, $a$ belongs to Flow A, starts at ingress $S_A$, exits at egress $D_A$ and passes through three congestion points in its path $\alpha_0$, $\alpha_1$ and $\alpha_2$; $x$ belongs to Flow X, starts at ingress $S_X$, exits at egress $D_X$ and passes through three congestion points in its path $\alpha_0$, $\alpha_3$ and $\alpha_4$; and so on.

The two critical packets we care about in this example are $a$ and $x$, which interact with each-other at their first congestion point $\alpha_0$, being scheduled by $\alpha_0$ at different times in the two cases ($a$ before $x$ in Case 1 and $x$ before $a$ in Case 2). But, notice that for both cases,
\begin{itemize}
\item $a$ enters the network from its ingress $S_A$ at congestion point $\alpha_0$ at time 0, and passes through two other congestion points $\alpha_1$ and $\alpha_2$ before exiting the network at time $(4 + 1)$. \footnote{$+1$ is added to indicate transmission time at the last congestion point. As mentioned before, we assume the propagation delay to the egress and the transmission time at the egress are both 0.}
\item $x$ enters the network from its ingress $S_X$ at congestion point $\alpha_0$ at time 0, and passes through two other congestion points $\alpha_3$ and $\alpha_4$ before exiting the network at time $(3 + 1)$.
\end{itemize}

$a$ interacts with packets from Flow C at its third congestion point $\alpha_2$, while $x$ interacts with a packet from Flow Z at its third congestion point $\alpha_4$. For both cases, 
\begin{itemize}
\item Two packets of Flow C ($c_1, c_2$) enter the network at times 2 and 3 at $\alpha_2$ before they exit the network at time $(2+1)$ and $(3+1)$ respectively.
\item $z$ enters the network at time 2 at $\alpha_4$ before exiting at time $2+1$. 
\end{itemize}

The difference between the two cases comes from how $a$ interacts with packets from Flow B at its second congestion point $\alpha_1$ and how $x$ interacts with packets from Flow Y at its second congestion points $\alpha_3$. Note that $\alpha_1$ and $\alpha_3$ are the last congestion points for Flow B and Flow Y packets respectively and their exit times from these congestion points directly determine their exit times from the network. 
\begin{itemize}
\item Three packets of Flow B ($b_1, b_2, b_3$) enter the network at times 2, 3 and 4 respectively at $\alpha_1$. In Case 1, they leave $\alpha_1$ at times $(2+1), (3+1), (4+1)$ respectively, providing no lee-way \footnote{It is required that $\alpha_1$ must schedule $a$ by at most time $3$ in order for it to exit the network at its target output time.} for $a$ at $\alpha_0$, which leaves $\alpha_1$ at time $(1+1)$. In Case 2, ($b_1, b_2, b_3$) leave at times $(3+1), (4+1), (5+1)$ respectively, providing lee-way for $a$ at $\alpha_0$, which leaves $\alpha_1$ at time $(2+1)$.
\item Two packets of Flow Y ($y_1, y_2$) enter the network at times 2 and 3 respectively at $\alpha_3$.  In Case 1, they leave at times $(3+1), (4+1)$ respectively, providing a lee-way for $x$ at $\alpha_0$, which leaves $\alpha_3$ at time $(2+1)$. In Case 2, ($y_1, y_2$) exit at times $(2+1), (3+1)$, providing no lee-way for $x$ at $\alpha_0$, which leaves $\alpha_3$ at time $(1+1)$. 
\end{itemize}

Note that the interaction of $a$ and $x$ with Flow C and Flow Z at their third congestion points respectively, is what ensures that their eventual exit time remains the same across the two cases inspite of the differences in how $a$ and $x$ are scheduled in their previous two hops.

Thus, we can see that $i(a)$, $o(a)$, $i(x)$, $o(x)$ are the same in both cases (also indicated in bold blue). Yet, \emph{Case 1} requires $a$ to be scheduled before $x$ at $\alpha_0$, else packets will get delayed at $\alpha_1$, since it is required that $\alpha_1$ schedules $a$ at a time no more than 3 units if it is to meet its target output time. \emph{Case 2} requires $x$ to be scheduled before $a$ at $\alpha_0$, else packets will be delayed at $\alpha_3$, where it is required to schedule $x$ at a time no more than 2 units if it is to meet its target output time. Since the attributes $(i(\cdot), o(\cdot), path(\cdot))$ for both $a$ and $x$ are exactly the same in both cases, any deterministic UPS with Blackbox Initialization will produce the same order for the two packets at $\alpha_0$, which contradicts the situation where we want $a$ before $x$ in one case and $x$ before $a$ in another.

\section{Deriving the Slack Equation}
\label{app:slackeqn}

We now prove that for any packet $p$ waiting at any node $\alpha$ at time $t_{now}$, the remaining slack of the last bit of $p$ is given by $slack(p,\alpha,t_{now}) = o(p) - t_{now} - t_{min}(p,\alpha,dest(p)) + T(p, \alpha)$.

Let $t_{wait}(p, \alpha, t_{now})$ denote the total time spent by $p$ on waiting behind other packets at the nodes in its path from $src(p)$ to $\alpha$ (including these two nodes) until time $t_{now}$. We define $t_{wait}(p, \alpha, t_{now})$, such that it excludes the transmission times at previous nodes which gets captured in $t_{min}$, but includes the local service time received by the packet so far at $\alpha$ itself.

\begin{subequations}
\small
\begin{align}
slack(p, \alpha, t_{now}) =& slack(p) - t_{wait}(p,\alpha, t_{now}) + T(p,\alpha) \label{subeq:remainingslack} \\
=& o(p) - i(p) - t_{min}(p, src(p), dest(p)) \nonumber \\
& - t_{wait}(p,\alpha, t_{now}) + T(p,\alpha)  \\
=& o(p) - i(p) - (t_{min}(p, src(p),\alpha) \nonumber \\
&+ t_{min}(p, \alpha, dest(p)) - T(p, \alpha)) \nonumber \\ 
& - t_{wait}(p,\alpha, t_{now}) + T(p,\alpha) \label{subeq:breaktmin} \\
=& o(p) - t_{min}(p, \alpha, dest(p)) + T(p, \alpha)  \nonumber \\
& - (i(p) + t_{min}(p, src(p),\alpha) \nonumber \\
& - T(p,\alpha) + t_{wait}(p,\alpha, t_{now})) \\
=& o(p) - t_{min}(p, \alpha, dest(p)) + T(p, \alpha) - t_{now}  \label{subeq:definetnow} 
\end{align}
\end{subequations}

(\ref{subeq:remainingslack}) is straightforward from our definition of LSTF and how the slack gets updated at every time slice. $T(p,\alpha)$ is added since $\alpha$ needs to locally consider the slack of the last bit of the packet in a store-and-forward network.
(\ref{subeq:breaktmin}) then uses the fact that for any $\alpha$ in $path(p)$,  $(t_{min}(p, src(p), dest(p)) = t_{min}(p, src(p),\alpha) + t_{min}(p, \alpha, dest(p)) - T(p,\alpha))$.  $T(p, \alpha)$ is subtracted here as it is accounted for twice when we break up the equation for $t_{min}(p, src(p), dest(p))$. (\ref{subeq:definetnow}) then follows from the fact that the difference between $t_{now}$ and $i(p)$ is equal to the total amount of time the packet has spent in the network until time $t_{now}$ i.e. $(t_{now} - i(p) = (t_{min}(p, src(p),\alpha) - T(p,\alpha)) + t_{wait}(p,\alpha, t_{now}))$. We need to subtract $T(p,\alpha)$, since by our definition, $t_{min}(p, src(p), \alpha)$ includes transmission time of the packet at $\alpha$.

\section{LSTF and EDF Equivalence}
\label{app:edf-lstf-equivalence}

In our network-wide extension of EDF scheduling, every router computes a \emph{local} deadline of a packet $p$ based on the static header value $o(p)$ and additional state information about the minimum time the packet would take to reach its destination from the router. More precisely, each router (say $\alpha$), uses $priority(p) = (o(p) - t_{min}(p, \alpha, dest(p)) + T(p,\alpha))$ to do priority scheduling, with $o(p)$ being the value carried by the packet header, initialized at the ingress and remaining unchanged throughout. EDF is equivalent to LSTF, in that for a given viable schedule, the two produce exactly the same replay schedule.

\paragraphb{Proof:} Consider any node $\alpha$ with non-empty queue at any given time $t_{now}$. Let $P(\alpha,t_{now})$ be the set of packets waiting at $\alpha$ at time $t_{now}$. A packet will then be scheduled by $\alpha$ as follows:

\paragraphi{With EDF:} 
Schedule packet $p_{edf}(\alpha, t_{now})$, where
\begingroup
\small
\begin{align*}
p_{edf}(\alpha, t_{now}) =& \argmin\limits_{p \in P(\alpha,t_{now})}(priority(p,\alpha)) \\
priority(p,\alpha) =& o(p) - t_{min}(p, \alpha, dest(p)) + T(p, \alpha)
\end{align*}
\endgroup

\paragraphi{With LSTF:} Schedule packet $p_{lstf}(\alpha, t_{now})$, where
\begingroup
\small
\begin{align*}
p_{lstf}(\alpha, t_{now}) =& \argmin\limits_{p \in P(\alpha,t_{now})}(slack(p,\alpha,t_{now})) \\
slack(p,\alpha,t_{now}) =& o(p) - t_{min}(p,\alpha,dest(p)) + T(p, \alpha) - t_{now}
\end{align*}
\endgroup

The above expression for $slack(p,\alpha,t_{now})$ has been derived in \S\ref{app:slackeqn}.
Thus, $slack(p,\alpha,t_{now}) = priority(p,\alpha) - t_{now}$. Since $t_{now}$ is the same for all packets, we can conclude that:

\begin{align*}
\argmin\limits_{p \in P(\alpha,t_{now})}(slack(p,\alpha,t_{now})) &= \argmin\limits_{p \in P(\alpha,t_{now})}(priority(p,\alpha)) \\
\implies p_{lstf}(\alpha, t_{now}) &= p_{edf}(\alpha, t_{now}) 
\end{align*}

Therefore, at any given point of time, all nodes with non-empty queues will schedule the same packet with both EDF and LSTF. \footnote{Assuming ties are broken in the same way for both per-router EDF and LSTF, such as by using FCFS.} Hence, EDF and LSTF are equivalent.

\section{Simple Priorities Replay Failure for Two Congestion Points Per Packet}
\label{app:prioritiesfailure}

\begin{figure}

\centering
\begin{subfigure}
\centering
\includegraphics[width=0.4\textwidth]{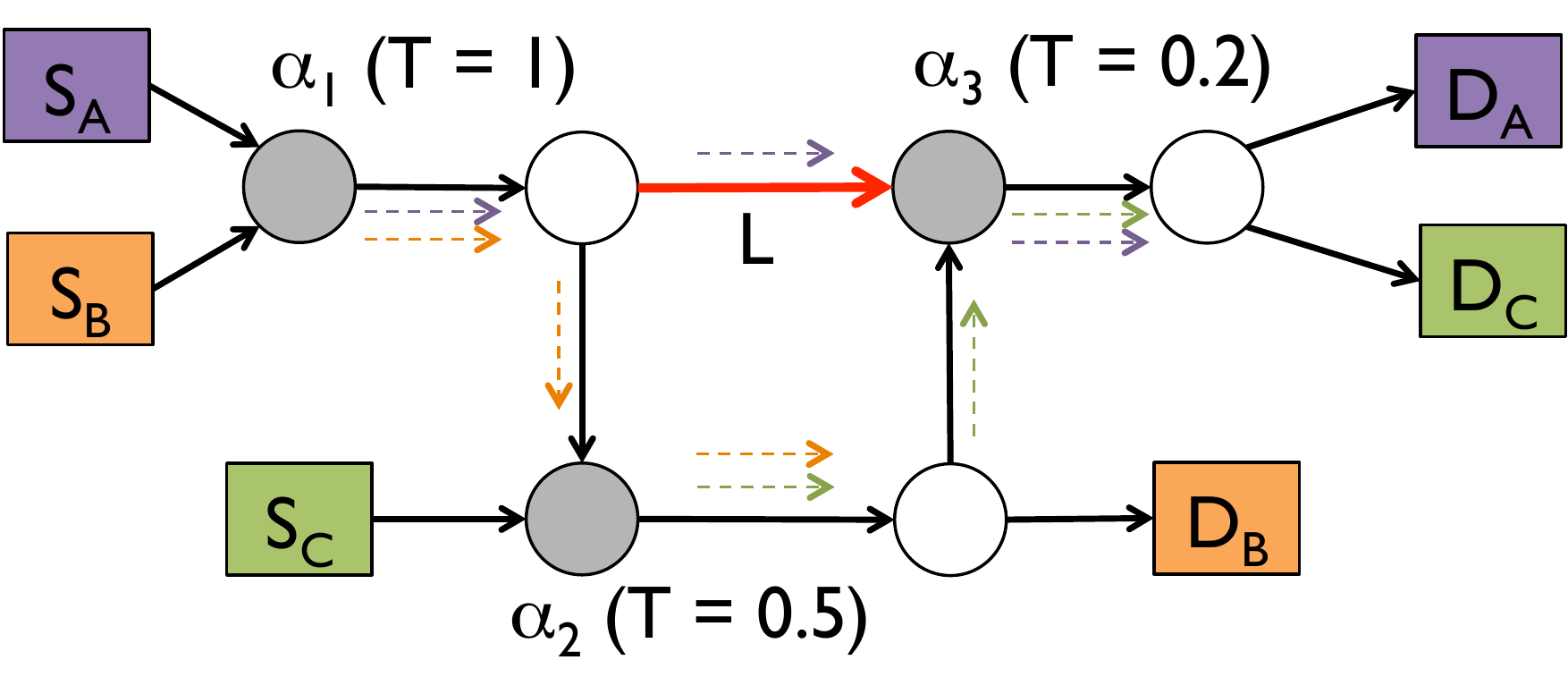}
\end{subfigure}

\centering
\small
\begin{tabular}[b]{|cc|}
\hline
{\bf Node} & {\bf Packet(arrival time, scheduling time)} \\
\hline
$\alpha_1$ & $a(0, 0), b(0, 1)$ \\
$\alpha_2$ & $b(2, 2), c(2, 2.5)$ \\
$\alpha_3$ & $c(3, 3), a(3, 3.2)$ \\
\hline
\multicolumn{2}{c}{}  \\
\multicolumn{2}{c}{}  \\
\end{tabular}

\caption{Example showing replay failure with simple priorities for a schedule with two congestion points per packet. A packet represented by $p$ belongs to flow $P$, with ingress $S_P$ and egress $D_P$, where $P \in \{A, B, C\}$. All packets are of the same size. For simplicity assume all links (except L) have a propagation delay of zero. L has a propagation delay of $2$. All uncongested routers (white circles), ingresses and egresses have a transmission time of zero. The three congestion points -- $\alpha_1, \alpha_2, \alpha_3$ have transmission times of $T = 1$ unit, $T = 0.5$ units and $T = 0.2$ units respectively.}
\label{fig:priorities2cpperflow}
\end{figure}

In Figure~\ref{fig:priorities2cpperflow}, we present an example which shows that simple priorities can fail in replay when there are two congestion points per packet, no matter what information is used to assign priorities. At $\alpha_1$, we need to have $priority(a) < priority(b)$, at $\alpha_2$ we need to have  $priority(b) < priority(c)$ and at $\alpha_3$ we need to have $priority(c) < priority(a)$. This creates a priority cycle where we need $priority(a) < priority(b) < priority(c) < priority(a)$, which can never be possible to achieve with simple priorities. 

We would also like to point out here that priority assignment for perfect replay in networks with smaller complexity (with single congestion point per packet) requires detailed knowledge about the topology and input load. More precisely, if a packet $p$ passes through congestion point $\alpha_p$, then its priority needs to be assigned as $priority(p) = o(p) - t_{min}(p, \alpha_p, dest(p)) + T(p, \alpha_p)$.\footnote{The proof that this would work for at most one congestion point per packet follows from the fact that the only scheduling decision made in a packet $p$'s path is at the single congestion point $\alpha_p$. This decision is same as what will be made with per-router EDF (just for at most one congestion point per packet), which we proved is equivalent to LSTF in \S\ref{app:edf-lstf-equivalence}, which in turn always gives a perfect replay for one (or to be more precise, at most two) congestion points per packet (as we shall prove in \S\ref{app:lstfproof}).} Hence, we need to know where the congestion point occurs in a packet's path, along with the final output times, to assign the priorities. In the absence of this knowledge, priorities cannot replay even a single congestion point. \eat{LSTF, on the other hand requires only the minimum taken to traverse a packet's path (from source to destination) and the packet's final output time to replay any schedule with at most two congestion points per packet (as we prove in \S\ref{app:lstfproof}).}

\balance
\section{LSTF: Perfect Replay for at most Two Congestion Points per Packet}
\label{app:lstfproof}

\subsection{Main Proof}

We first prove that LSTF can replay any schedule with at most two congestion points per packet. Note that we work with bits in our proof, since we assume a pre-emptive version of LSTF. Due to store-and-forward routers, the remaining slack of a packet at a particular router is represented by the slack of the last bit of the packet (with all other bits of the packet having the same slack as the last bit). 

In order for a replay failure to occur, there must be at least one overdue packet, where a packet $p$ is said to be overdue if $o'(p) > o(p)$. This implies that $p$ must have spent all of its slack while waiting behind other packets at a queue in some node $\alpha$ at say time $t$, such that $slack(p, \alpha, t) < 0$. Obviously, $\alpha$ must be a congestion point.

\paragraphb{Necessary Condition for Replay Failure with LSTF:} If a packet $p^*$ sees negative slack at a congestion point $\alpha$ when its last bit exits $\alpha$ at time $t^*$ in the replay (i.e. $slack(p^*, \alpha, t^*) < 0$), then $(\exists p \in pass(\alpha) \mid i'(p, \alpha) \le t^* \text{ and } i'(p, \alpha) > o(p, \alpha))$. We prove this in \S\ref{app:keyproof}.

\paragraphb{Key Observation:} \emph{When there are at most two congestion points per packet, then no packet $p$ can arrive at any congestion point $\alpha$ in the replay, after its corresponding scheduling time at $\alpha$ in the original schedule (.i.e. $i'(p, \alpha) > o(p, \alpha)$ is not possible). Therefore, by the necessary condition above, no packet can see a negative slack at any congestion point.}

\paragraphb{Proof by contradiction:} 
Suppose that there exists $\alpha^*$, which is the first congestion point (in time) that sees a packet which arrives after its corresponding scheduling time in the original schedule. Let $p^*$ be this first packet that arrives after the corresponding scheduling time in the original schedule at $\alpha^*$ ($i'(p^*, \alpha^*) > o(p^*, \alpha^*)$).  
Since there are at most two congestion points per packet, either $\alpha^*$ is the first congestion point seen by $p^*$ or the last (or both). 

\paragraphi{(1)} If $\alpha^*$ is the first congestion point seen by $p^*$, then clearly, $i'(p^*, \alpha^*) = i(p^*, \alpha^*) \le o(p^*, \alpha^*)$. This contradicts our assumption that $i'(p^*, \alpha^*) > o(p^*, \alpha^*)$.

\paragraphi{(2)} If $\alpha^*$ is not the first congestion point seen by $p^*$, then it is the last congestion point seen by $p^*$. If $i'(p^*, \alpha^*) > o(p^*, \alpha^*)$, then it would imply that $p^*$ saw a negative slack before arriving at $\alpha^*$. Suppose $p^*$ saw a negative slack at a congestion point $\alpha_{prev}$, before arriving at $\alpha^*$ when its last bit exited $\alpha_{prev}$ at time $t_{prev}$. Clearly, $t_{prev} < i'(p^*, \alpha^*)$. As per our necessary condition, this would imply that there must be another packet $p'$, such that $i'(p', \alpha_{prev}) > o(p', \alpha_{prev})$ and $i'(p', \alpha_{prev}) \le t_{prev} < i'(p^*, \alpha^*)$. This contradicts our assumption that $\alpha^*$ is the first congestion point (in time) that sees a packet which arrives after its corresponding scheduling time in the original schedule.

Hence, no congestion point can see a packet that arrives after its corresponding scheduling time in the original schedule (and therefore no packet can get overdue) when there are at most two congestion points per packet.

We finally present, in \S\ref{app:lstfbad}, an example where LSTF replay failure occurs with no more than three congestion points per packet, thus completing our proof that LSTF can replay any schedule with at most two congestion points per flow and can fail beyond that.

\subsection{Proof for Necessary Condition for Replay Failure with LSTF}
\label{app:keyproof}

We start with the following observation that we use in our proof.

\paragraphb{Observation 1:} If all bits of a packet $p$ exit a router $\alpha$ by time $o(p, \alpha) + T(p, \alpha)$, then $p$ cannot see a negative slack at $\alpha$. 

\paragraphb{Proof for Observation 1:} As shown previously in \S\ref{app:slackeqn},
\begingroup
\small
\begin{align*}
&slack(p, \alpha, t) = o(p) - t_{min}(p, \alpha, dest(p)) + T(p, \alpha) - t \\
\end{align*}
\endgroup
Therefore,
\begingroup
\small
\begin{align*}
&slack(p, \alpha, o(p, \alpha) + T(p, \alpha)) \\
&= o(p) - t_{min}(p, \alpha, dest(p)) + T(p,\alpha) - (o(p, \alpha) + T(p, \alpha)) \\
&\text{But, } o(p) = o(p, \alpha) + t_{min}(p, \alpha, dest(p)) + wait(p, \alpha, dest(p)) \\ 
&\implies slack(p, \alpha, o(p, \alpha) + T(p, \alpha)) = wait(p, \alpha, dest(p)) \\
&\implies slack(p, \alpha, o(p, \alpha) + T(p, \alpha)) \ge 0
\end{align*}
\endgroup
where $wait(p, \alpha, dest(p))$ is the time spent by $p$ in waiting behind other packets in the original schedule, after it left $\alpha$, which is clearly non-negative.
\newline
\newline
\noindent We now move to the main proof for the necessary condition. 

\paragraph{Necessary Condition for Replay Failure:}  If a packet $p^*$ sees negative slack at a congestion point $\alpha$ when its last bit exits $\alpha$ at time $t^*$ in the replay (i.e. $slack(p^*, \alpha, t^*) < 0$), then $(\exists p \in pass(\alpha) \mid i'(p, \alpha) \le t^* \text{ and } i'(p, \alpha) > o(p, \alpha))$.

\paragraph{Proof by Contradiction:} Suppose this is not the case .i.e. there exists $p^*$ whose last bit exits $\alpha$ at time $t^*$, such that $slack(p^*, \alpha, t^*) < 0$ and $(\forall p \in pass(\alpha) \mid i'(p, \alpha) > t^* \text{ or } i'(p, \alpha) \le o(p, \alpha))$. In other words, if $i'(p,\alpha) \le t^*$, then $i'(p,\alpha) \le o(p,\alpha)$. We can show that if this holds, then $p^*$ cannot see a negative slack at $\alpha$, thus violating our assumption.

We take the set of all bits which exit $\alpha$ at or before time $t^*$ in the LSTF replay schedule. We denote this set as $S_{bits}(\alpha, t^*)$. Since all of these bits (and the corresponding packets) must arrive at or before time $t^*$, as per our assumption, $(\forall b \in S_{bits}(\alpha, t^*) \mid i'(p_b, \alpha) \le o(p_b, \alpha))$, where $p_b$ is denoted as the packet to which bit $b$ belongs. Note that $S_{bits}(\alpha, t^*)$ also includes all bits of $p^*$ as per our definition of $S_{bits}(\alpha, t^*)$. 

We now prove that no bit in $S_{bits}(\alpha, t^*)$ can see a negative slack (and therefore $p^*$ cannot see a negative slack at $\alpha$), leading to a contradiction. The proof comprises of two steps: 

\paragraphi{Step 1:} Using the same input arrival times of each packet at $\alpha$ as in the replay schedule, we first construct a \emph{feasible schedule} at $\alpha$ up until time $t^*$, denoted by $FS(\alpha, t^*)$, where by feasibility we mean that no bit in $S_{bits}(\alpha, t^*)$ sees a negative slack. 

\paragraphi{Step 2: } We then do an iterative transformation of $FS(\alpha, t^*)$ such that the bits in $S_{bits}(\alpha, t^*)$ are scheduled in the order of their \emph{least remaining slack times}. This reproduces the LSTF replay schedule from which $FS(\alpha, t^*)$ was constructed in the first place. However, while doing the transformation we show how the schedule remains feasible at every iteration, proving that the LSTF schedule finally obtained is also feasible up until time $t^*$. In other words, no packet sees a negative slack at $\alpha$ in the resulting LSTF replay schedule up until time $t^*$, contradicting our assumption that $p^*$ sees a negative slack when it exits $\alpha$ at time $t^*$ in the replay.

\noindent We now discuss these two steps in details.

\paragraphb{Step 1:} Construct a feasible schedule at $\alpha$ up until time $t^*$ (denoted as $FS(\alpha, t^*)$) for which no bit in $S_{bits}(\alpha, t^*)$ sees a negative slack.

\paragraphi{(i)} Algorithm for constructing $FS(\alpha, t^*)$: Use priorities to schedule each bit in $S_{bits}(\alpha, t^*)$, where $\forall b \in S_{bits}(\alpha, t^*) \mid priority(b) = o(p_b, \alpha)$.
(Note that since both $FS(\alpha, t^*)$ and LSTF are work-conserving, $FS(\alpha, t^*)$ is just a shuffle of the LSTF schedule up until $t^*$. The set of time slices at which a bit is scheduled in $FS(\alpha, t^*)$ and in the LSTF schedule up until $t^*$ remains the same, but \emph{which} bit gets scheduled at a given time slice is different.)

\paragraphi{(ii)} In $FS(\alpha, t^*)$, all bits $b$ in $S_{bits}(\alpha, t^*)$ exit $\alpha$ by time $o(p_b, \alpha) + T(p_b, \alpha)$.

\paragraphi{Proof by contradiction:} 
Suppose the statement is not true and consider the first bit $b^*$ that exits after time $(o(p_{b^*},\alpha) + T(p_{b^*},\alpha))$. We term this as $b^*$ got late at $\alpha$ due to $FS(\alpha, t^*)$. Remember that, as per our assumption, $(\forall b \in S_{bits}(\alpha, t^*) \mid i'(p_b, \alpha) \le o(p_b, \alpha))$. Thus, given that all bits of $p_{b^*}$ arrive at or before time $o(p_{b^*}, \alpha)$, the only reason why the delay can happen in our work-conserving $FS(\alpha, t^*)$ is if some other higher priority bits were being scheduled after time $o(p_{b^*}, \alpha)$, resulting in $p_{b^*}$ not being able to complete its transmission by time $(o(p_{b^*},\alpha) + T(p_{b^*}, \alpha))$. However, as per our priority assignment algorithm, any bit $b'$ having a higher priority than $b^*$ at $\alpha$ must have been scheduled before the first bit of $p_{b^*}$ in the non-preemptible original schedule, implying that $(o(p_{b'}, \alpha) + T(p_{b'}, \alpha)) \le o(p_{b^*}, \alpha)$. Therefore, a bit $b'$ being scheduled after time $o(p_{b^*},\alpha)$, implies it being scheduled after time $(o(p_{b'}, \alpha) + T(p_{b'}, \alpha))$. This contradicts our assumption that $b^*$ is the first bit to get late at $\alpha$ due to $FS(\alpha, t^*)$. Therefore, all bits $b$ in $S_{bits}(\alpha, t^*)$ exit $\alpha$ by time $o(p_b, \alpha) + T(p_b, \alpha)$ as per the schedule $FS(\alpha, t^*)$.

\paragraphi{(iii)}  Since all bits in $S_{bits}(\alpha, t^*)$ exit by time $o(p_b, \alpha) + T(p_b, \alpha)$ due to $FS(\alpha, t^*)$, no bit in $S_{bits}(\alpha, t^*)$ sees a negative slack at $\alpha$ (from Observation 1).

\paragraphb{Step 2: } \emph{Transform} $FS(\alpha, t^*)$ into a feasible LSTF schedule for the single switch $\alpha$ up until time $t^*$. 

(Note: The following proof is inspired from the standard LSTF optimality proof that shows that for a single switch, any feasible schedule can be transformed to an LSTF schedule). 

Let $fs(b, \alpha, t^*)$ be the scheduling time slice for bit $b$ in $FS(\alpha, t^*)$. The transformation to LSTF is carried out by the following pseudocode:

\vspace{10pt}
\begin{algorithmic}[1]
\While{true}
    \State Find two bits, $b_1$ and $b_2$, such that:\par
    \hskip\algorithmicindent $(fs(b_1, \alpha, t^*) < fs(b_2, \alpha, t^*))$ {\bf and} \par
    \hskip\algorithmicindent $(slack(b_2, \alpha, fs(b_1, \alpha, t^*))$ \par
    \hskip\algorithmicindent $< slack(b_1, \alpha, fs(b_1, \alpha, t^*)))$ {\bf and} \par
    \hskip\algorithmicindent $(i'(b_2, \alpha, t^*) \le fs(b_1, \alpha, t^*))$
    \If { no such $b_1$ and $b_2$ exist }
        \State $FS(\alpha, t^*)$ is an LSTF schedule
        \State {\bf break}
    \Else
        \State $swap(fs(b_1, \alpha, t^*), fs(b_2, \alpha, t^*))$ \label{swapstep} \Comment swap the scheduling times of the two bits.
         \footnote{Note that we are working with bits here for easy expressibility. In practice, such a swap is possible under the preemptive LSTF model.}
    \EndIf
\EndWhile
\State Shuffle the scheduling time of the bits belonging to the same packet, to ensure that they are in order.
\label{shufflestep1}
\State Shuffle the scheduling time of the same-slack bits such that they are in FIFO order
\label{shufflestep2}

\end{algorithmic}

Line~\ref{swapstep} above will not cause $b_1$ to have a negative slack, when it gets scheduled at $fs(b_2, \alpha, t^*)$ instead of $fs(b_1, \alpha, t^*)$. This is because the difference in $slack(b_2, \alpha, t)$ and $slack(b_1, \alpha, t)$ is independent of $t$ and so:
\begingroup
\small
\begin{align*}
& slack(b_2, \alpha, fs(b_1, \alpha, t^*)) < slack(b_1, \alpha, fs(b_1, \alpha, t^*)) \\
\implies & slack(b_2, \alpha, fs(b_2, \alpha, t^*)) < slack(b_1, \alpha, fs(b_2, \alpha, t^*))
\end{align*}
\endgroup
Since $FS(\alpha, t^*)$ is feasible before the swap, $slack(b_2, \alpha, fs(b_2, \alpha, t^*)) \ge 0$. Therefore, $slack(b_1, \alpha, fs(b_2, \alpha, t^*)) > 0$ and the resulting $FS(\alpha, t^*)$ after the swap remains feasible.

Lines \ref{shufflestep1} and \ref{shufflestep2} will also not result in any bit getting a negative slack, because all bits participating in the shuffle have the same slack at any fixed point of time in $\alpha$.

Therefore, no bit in $S_{bits}(\alpha, t^*)$ has a negative slack at $\alpha$ after any iteration.

Since no bit in $S_{bits}(\alpha, t^*)$ has a negative slack at $\alpha$ in the swapped LSTF schedule, it contradicts our statement that $p^*$ sees a negative slack when its last bit exits $\alpha$ at time $t^*$. Hence proved that if a packet $p^*$ sees a negative slack at congestion point $\alpha$ when its last bit exits $\alpha$ at time $t^*$ in the replay, then there must be at least one packet that arrives at $\alpha$ in the replay at or before time $t^*$ and later than the time at which it is scheduled by $\alpha$ in the original schedule.  

\subsection{Replay Failure Example with LSTF}
\label{app:lstfbad}

\begin{figure}

\centering
\begin{subfigure}
\centering
\includegraphics[width=0.45\textwidth]{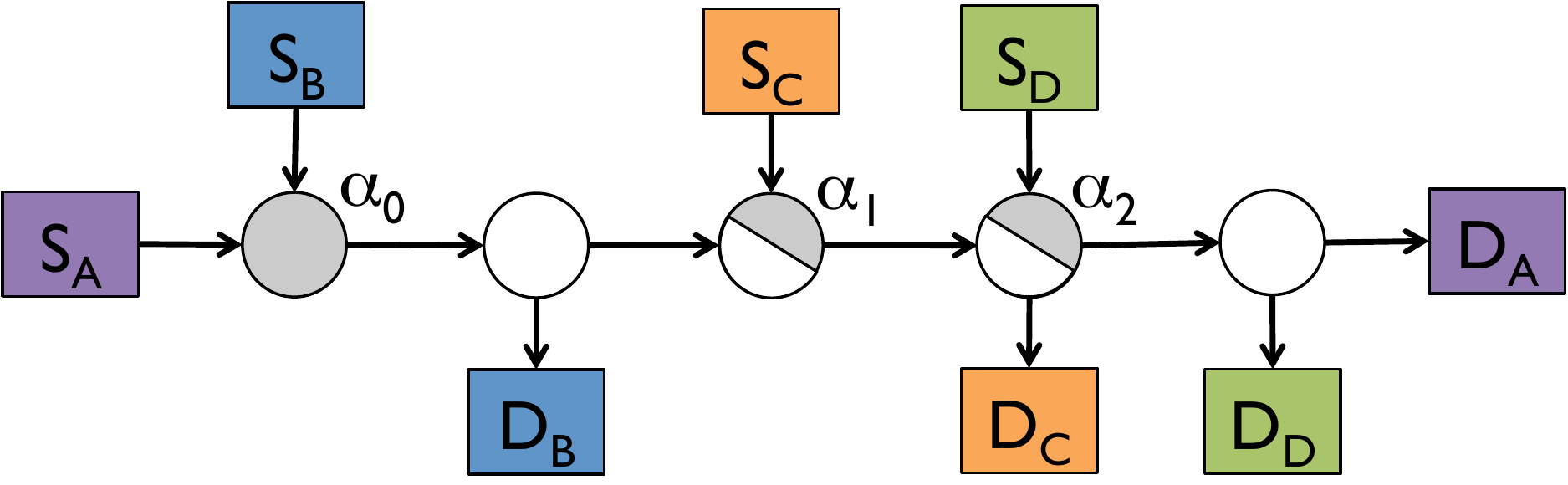}
\end{subfigure}

\centering
\small
\begin{tabular}[b]{|cc|}
\hline
\multicolumn{2}{|c|}{\bf Original Schedule} \\
\hline
{\em Node} & {\em Packet(arrival time, scheduling time)} \\
$\alpha_0$ & $a(0, 0), b(0, 1)$ \\
$\alpha_1$ & $a(1, 1), c_1(2, 2), c_2(3, 3)$ \\
$\alpha_2$ & $d_1(2, 2), d_2(3, 3), a(2, 4)$ \\
\hline
\multicolumn{2}{|c|}{\bf LSTF Replay} \\
\hline
{\em Node} & {\em Packet(arrival time, scheduling time)} \\
$\alpha_0$ & $b(0, 0), a(0, 1)$ \\
$\alpha_1$ & $c_1(2, 2), a(2, 3),$ \textcolor{red}{$c_2(3, 4)$} \\
$\alpha_2$ & $d_1(2, 2), d_2(3, 3), a(4, 4)$ \\
\hline
\end{tabular}

\caption{Example showing replay failure with LSTF when there is a flow with three congestion points. A packet represented by $p$ belongs to flow $P$, with ingress $S_P$ and egress $D_P$, where $P \in \{A, B, C, D\}$. For simplicity assume all links have a propagation delay of zero. All uncongested routers (white), ingresses and egresses have a transmission time of zero. The three congestion points (shaded grey) have transmission times of $T = 1$ unit}
\label{fig:lstfbad}

\end{figure}

In Figure~\ref{fig:lstfbad}, we present an example where a flow passes through three congestion points and a replay failure occurs with LSTF. When packet $a$ arrives at $\alpha_0$, it has a slack of 2 (since it waits behind $d_1$ and $d_2$ at $\alpha_2$), while at the same time, packet $b$ has a slack of 1 (since it waits behind $a$ at $\alpha_0$). As a result, $b$ gets scheduled before $a$ in the LSTF replay. $a$ therefore arrives at $\alpha_1$ with slack 1 at time $2$. $c_1$ with a zero slack is prioritized over $a$. This reduces $a$'s slack to zero at time $3$, when $c_2$ is also present at $\alpha_1$  with zero slack. Scheduling $a$ before $c_2$, will result in $c_2$ being overdue (as shown). Likewise, scheduling $c_2$ before $a$ would have resulted in $a$ getting overdue. Note that in this failure case, $a$ arrives at $\alpha_1$ at time $2$, which is greater than $o(a, \alpha_1) = 1$.

\end{appendices}

\end{document}